\documentclass[lettersize,journal]{IEEEtran}
\usepackage{amsmath,amsfonts}
\usepackage{array}
\usepackage[caption=false,font=normalsize,labelfont=footnotesize,textfont=footnotesize]{subfig}
\usepackage{textcomp}
\usepackage{stfloats}
\usepackage{url}
\usepackage{verbatim}
\usepackage{graphicx}
\usepackage{cite}
\usepackage{threeparttable}
\usepackage{booktabs}
\usepackage{makecell}
\usepackage{adjustbox}

\usepackage[colorlinks,linkcolor=green,anchorcolor=blue,citecolor=green]{hyperref}
\usepackage[ruled, vlined, linesnumbered]{algorithm2e}

\usepackage{listings}
\usepackage{xcolor}
\usepackage{rotating}
\usepackage{multirow}
\definecolor{codegreen}{rgb}{0,0.6,0}
\definecolor{codegray}{rgb}{0.5,0.5,0.5}
\definecolor{codepurple}{rgb}{0.58,0,0.82}
\definecolor{backcolour}{rgb}{0.95,0.95,0.92}

\lstdefinestyle{mystyle}{
    backgroundcolor=\color{backcolour},   
    commentstyle=\color{codegreen},
    keywordstyle=\color{magenta},
    numberstyle=\tiny\color{codegray},
    stringstyle=\color{codepurple},
    basicstyle=\ttfamily\footnotesize,
    breakatwhitespace=false,         
    breaklines=true,                 
    captionpos=b,                    
    keepspaces=true,                 
    numbers=left,                    
    numbersep=5pt,                  
    showspaces=false,                
    showstringspaces=false,
    showtabs=false,                  
    tabsize=2,
    xleftmargin=1em
}
\usepackage{pifont}

\usepackage{xspace}
\newcommand{\mypara}[1]{\smallskip\noindent\textbf{#1.} \xspace}

\begin{document}

\title{PCG: Mitigating Conflict-based Cache Side-channel Attacks with Prefetching}

\author{Fang Jiang,~\IEEEmembership{Student Member,~IEEE}, Fei Tong,~\IEEEmembership{Member,~IEEE}, Hongyu Wang, Xiaoyu Cheng, Zhe Zhou, Ming Ling,~\IEEEmembership{Member,~IEEE}, and Yuxing Mao
        
        \IEEEcompsocitemizethanks{\IEEEcompsocthanksitem This work is supported in part by the National Natural Science Foundation of China (No. 61971131), and in part by ``Zhishan'' Scholars Programs of Southeast University. (Corresponding Author: Fei Tong)
			\IEEEcompsocthanksitem F. Jiang, F. Tong, X. Cheng and Z. Zhou are with the School of Cyber Science and Engineering, Southeast University, Nanjing, Jiangsu, China (\{fangjiangff, ftong, xiaoyu\_cheng, 220215213\}@seu.edu.cn).
			\IEEEcompsocthanksitem F. Tong is also with Jiangsu Province Engineering Research Center of Security for Ubiquitous Network, China, and the Purple Mountain Laboratories, Nanjing, Jiangsu, China.
			\IEEEcompsocthanksitem H. Wang is with the State Key Laboratory of Power Transmission Equipment \& System Security and New Technology, Chongqing University, and Wiscom System Co., LTD (awang@wiscom.com.cn).
			\IEEEcompsocthanksitem M. Ling is with the School of Integrated Circuits, Southeast University, Nanjing, Jiangsu, China (trio@seu.edu.cn).
			\IEEEcompsocthanksitem Y. Mao is with the State Key Laboratory of Power Transmission Equipment \& System Security and New Technology, Chongqing University (myx@cqu.edu.cn).
		}
}


\maketitle
\begin{abstract}
To defend against conflict-based cache side-channel attacks, cache partitioning or remapping techniques were proposed to prevent set conflicts between different security domains or obfuscate the locations of such conflicts. But such techniques complicate cache design and may result in significant performance penalties. Therefore, there have been lightweight prefetching-based schemes proposed to introduce noise to confuse attackers’ observation. However, we have validated experimentally that relying on prefetching to only introduce noise is insufficient, as attackers can still reliably distinguish the victim's cache accesses. This paper proposes a novel prefetching-based scheme, called PCG. It combines adding victim-irrelevant cache occupancy changes and reducing victim-relevant cache occupancy changes to disrupt attackers by generating noisy and indistinguishable cache access patterns. Additionally, PCG can either work independently or seamlessly be integrated with most of the commonly used prefetchers. We have implemented and evaluated PCG in both gem5 and the open-source RISC-V core BOOMv3. The evaluation results show the PCG's robust security superior to the existing solutions, while without resulting in significant performance degradation. According to the evaluation based on the SPEC CPU 2017 benchmark suite, PCG even shows an average performance improvement of about 1.64\%. Moreover, it incurs only 1.26\% overhead on hardware resource consumption.

\end{abstract}


\section{Introduction}

Modern processors have introduced many optimization mechanisms, such as data prefetching, speculative execution, and data caching. While these technologies provide performance improvements, they also lead to prominent security issues at the micro-architecture level~\cite{LeakingSecretsModern2022,FLUSHRELOADhigh2014,MemoryAwareDenialofServiceAttacks2022}. Cache provides a large attack surface among all micro-architectures~\cite{LastLevelCacheSideChannel2015,Cachetemplateattacks2015,ARMageddonCacheAttacks2016, SideChannelAttacks2021}. Attackers use the time difference between cache hits and misses to launch attacks across different processes, platforms, and virtual machines. With the disclosure of new attacks such as Spectre~\cite{SpectreAttacksExploiting2019} and Meltdown~\cite{meltdown2020}, which exploit transient executions and cache features, the threat of cache side-channel attacks has further expanded. 

Among existing cache side-channel attacks, the conflict-based attack is one of the most practical and powerful ones, which leverages cache set conflicts to trigger cache evictions for launching attacks. Previous security-enhanced cache structures~\cite{CATalyst2016,PLCache2007,Newcachedesigns2007, MIRAGE2021}, utilizing cache partitioning or remapping techniques, have been shown to be effective in mitigating conflict-based attacks. However, these methods significantly complicate the design of caches and may introduce substantial overhead in terms of both performance and hardware resources. 

There have been some studies proposing prefetching-based mechanisms, such as Disruptive Prefetching (DP)~\cite{Disruptive2015} and PREFENDER~\cite{PREFENDER2022}, to defend against cache side-channel attacks. Such mechanisms do not target the design of secure cache architecture. Instead, they are proposed based on the fact that hardware prefetchers speculatively bring data/instructions into the cache, including those that have never actually been used by the victim, thereby which can be utilized as noise to confuse the attacker's observation of the victim's cache access pattern. Prefetching-based schemes enable the design of attack mitigation with fewer hardware modifications and smaller performance penalties. 

However, DP~\cite{Disruptive2015} causes cache pollution because it always triggers a large number of prefetches regardless of whether an attack occurs. PREFENDER~\cite{PREFENDER2022} triggers prefetching only when it detects the cache access times of {\tt load} instructions exceeding a fixed threshold. This leads to a situation where PREFENDER can be easily bypassed if the attacker uses multiple {\tt load} instructions and each instruction has fewer cache accesses than the threshold. Furthermore, both DP and PREFENDER do not reduce victim-relevant cache footprints, which are the cache occupancy changes caused by the victim's installation and eviction of cache lines. This still allows the attacker to distinguish the victim's cache footprints from the added noise by repeating the attack enough times~\cite{ASecurityRISC2023}.

To this end, this paper proposes a scheme called prefetching-based cache guard (PCG) to mitigate conflict-based cache side-channel attacks. PCG confuses the attacker’s observation of the victim’s cache access pattern by adding victim-irrelevant cache footprints as interfering noise and reducing victim-relevant cache footprints. As a result, the attacker cannot efficiently distinguish the left victim-relevant cache footprints from the added noise, even after repeating the attack numerous times. Specifically, to reduce victim-relevant cache footprints, PCG not only makes potential victim lines are evicted from the cache as early as possible by assigning them the highest replacement priority, but also prefetches these lines back into the cache if they may have been evicted due to the victim's activity. To add victim-irrelevant cache footprints, PCG initiates random forward or backward prefetches when there are cache misses and adjusts the prefetching addresses based on the program's cache set accesses.

The main contributions of this paper are as follows:
\begin{itemize}
    \item We propose PCG, a lightweight solution for mitigating conflict-based cache side-channel attacks. It works effectively even when the attacks are launched repeatedly a large number of times, thus demonstrating robust security that is superior to existing similar solutions.

    \item We show and verify that conflict-based cache side-channel attacks lead to a large number of MSHR (Miss Status Handling Register) misses (see Section~\ref{section:Unsafe}), which is utilized to efficiently add (reduce) the victim-irrelevant (victim-relevant) cache footprints in PCG.

    \item{We demonstrate experimentally how to break previous prefetching-based defenses,  PREFENDER~\cite{PREFENDER2022} and DP~\cite{Disruptive2015}, and discuss the reasons why and the scenarios where they are not effective in defending against conflict-based cache side-channel attacks (see Section~\ref{section:Security}).}

    \item{We have implemented and evaluated PCG using both gem5 and the open-source RISC-V core, BOOMv3~\cite{SonicBOOM3rdGeneration2020}, in the register transfer level (RTL). The evaluation results based on the SPEC CPU 2017 benchmark suite (SPEC17)~\cite{SPECCPU2017} show an average performance improvement of 1.64\% when PCG is enabled. The on-board hardware consumption increases by only about 1.26\% (see Section~\ref{section:Evaluation}).}
\end{itemize}

\section{Background}
\label{section:Background}

\subsection{Cache and MSHR}
Caches are designed to reduce the significant gap in speed between CPU process and memory access. The cache hierarchy is usually categorized into three levels, including the first-level cache (L1 cache), the second-level cache (L2 cache), and the Last-level cache (L3 cache/LLC). L1 and L2 caches are typically core-private, while LLC is shared among multiple cores.

Modern processors commonly adopt a set-associative cache design. Every memory address is divided into three parts, including {\it tag}, {\it index}, and {\it offset}. The cache primarily consists of a tag array and a data array. The {\it index} of an address identifies a cache set. Then by comparing the tag part of the address with each of the tags stored in the tag array of the identified set, a cache hit occurs if a match is found. Such a design result in memory addresses with the same {\it index} all being mapped to the same set, speeding up the lookup. However, cache set conflicts may happen, since the capacity of a set is limited, which can be exploited by conflict-based cache side-channel attackers to evict victim cache lines.

MSHR is an indispensable component for designing non-blocking caches in modern high-performance processors. It is used to manage the state of memory requests that result in cache misses, by merging multiple requests for the same cache line into a single entry. An MSHR miss indicates the need to allocate a new entry to handle the first miss for a specific cache line. For example, when the core accesses {\tt 0x8000\_a004} and {\tt 0x8000\_a0008} for the first time, both corresponding load instructions result in cache misses. However, as two request addresses belong to the same memory block (addresses differ only within the block offset), they will use the same MSHR entry to handle the cache miss, and thus there is only one MSHR miss.

\subsection{Conflict-based Cache Side-channel Attacks}
\label{subsection:cacheSideAttacks}

Conflict-based cache side-channel attacks exploit the fact that set conflicts exist in regular set-associative caches~\cite{MIRAGE2021}. The attacker occupies cache sets to evict the cache lines of the victim by constructing set conflicts. Evict+Reload,Evict+Time, and Prime+Probe are typically patterns utilized in conflict-based attacks.

\textbf{Evict+Reload:} The attacker initially accesses a large number of addresses which are mapped to the same cache set as the target address (i.e., these addresses contain the same {\it index} bits as the target address, the set of which is referred to as an eviction set (ES)), to achieve the effects similar to {\tt clflush} (Phase 1). After a certain interval, the attacker probes the target address again (Phase 3), and if a cache hit occurs, it can be deduced that the victim accessed this line in Phase 2.

 {\textbf{Evict+Time:} In Phase 1, this attack builds and accesses ES in order to evict the target cache lines, which contain the victim's sensitive data. In Phase 3, the attacker monitors the victim's overall execution time. A cache miss happens if the victim utilizes the target line, resulting in a longer execution time for an entire security-critical operation.}

\textbf{Prime+Probe:} The attacker adopting the Prime+Probe pattern accesses the addresses in the ES to fill target cache set(s) (Phase 1), waits for the victim to perform an operation (Phase 2), and then re-accesses the addresses in the ES (Phase 3). A cache miss indicates that the victim accessed the target address in Phase 2.

\subsection{Hardware Prefetcher}
Hardware Prefetchers request the memory blocks to be stored in the cache in advance according to the memory access history, thereby increasing the cache hit rate. One of the important properties of the prefetcher is its aggressiveness, which can be measured by the prefetching degree that specifies how many memory blocks are prefetched at a time. 

Various hardware data prefetchers~\cite{BingoSpatialData2019,BouquetInstructionPointers2020,BertiAccurateLocalDelta2022,PuppeteerRandomForest2022} are currently available, differing from each other primarily in terms of timing and prefetching strategies. Commonly used prefetchers include the Next-Line prefetcher and the Stride prefetcher. The main idea of the former is to always prefetch the next memory block after the currently accessed block. Prefetching is triggered in a Stride prefetcher when a fixed step exists for continuously accessing a number of memory blocks, which is very effective for accessing a dense data matrix and array. Arrays are often accessed in programs by loops with a fixed step size. For example, when accessing the $i^\mathrm{th}$, $(i+10)^\mathrm{th}$ and $(i+20)^\mathrm{th}$ elements of an array, the prefetcher learns a stride of 10 and begins prefetching the $(i+30)^\mathrm{th}$ and $(i+40)^\mathrm{th}$ elements into cache.

\section{MOTIVATION AND OBSERVATIONS}
\label{section:Unsafe}

\textbf{Observation 1: The typical confilct-based attacks will result in consecutive MSHR misses.} In conflict-based cache side-channel attacks, attackers initialize cache status by accessing the addresses in ES. Suppose that ES contains $N\times{}W$ addresses, where $W$ refers to the associativity of the cache. Considering a random replacement policy, $N$ needs to be large enough to ensure that the entire cache set is evicted. Typically, in the implementation of the attack, $N$ is set to 4 in L1 DCache, which results in a success rate of 99\% for evicting a specific cache line~\cite{Gonzalez2019ReplicatingAM}. The addresses accessed by the attacker from the ES belong to different memory blocks, which require the allocation of different entries for processing, leading to consecutive MSHR misses.

\begin{figure}[!t]
  \centering
  \includegraphics[width=\linewidth]{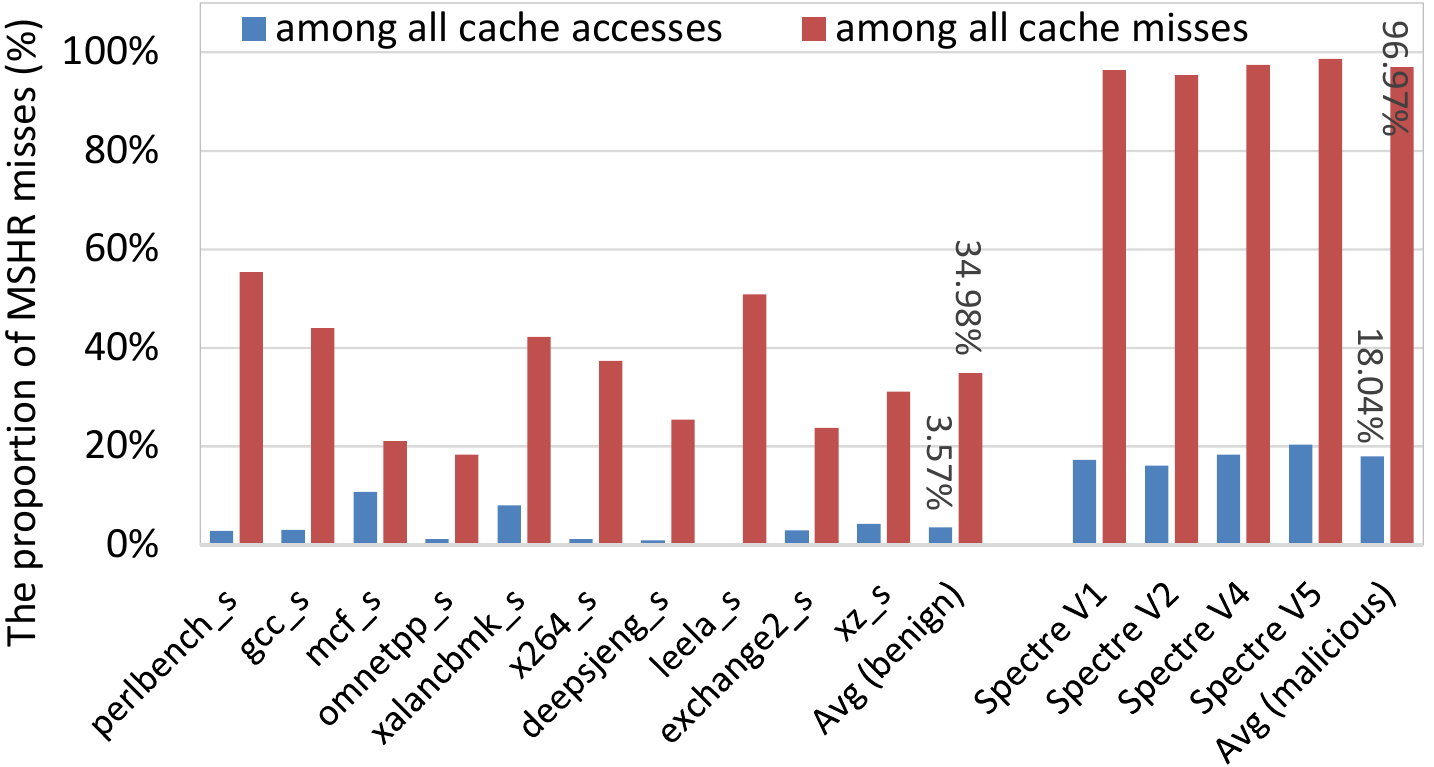}
  \caption{The proportion of MSHR misses in total cache accesses and cache misses.}
  \label{dal-ration}
\end{figure}

To verify that there are a significant number of MSHR misses during the conflict-based cache side-channel attacks, we have run the SPEC17 benchmarks and the Spectre-type Evict+Reload attacks (including Spectre V1, V2, V4, and V5) in gem5, representing benign and attack programs, respectively. For the first 10 million instructions of each program, we count the number of cache accesses, the number of cache misses, and the number of MSHR misses. The proportions of MSHR misses among all cache accesses and those among all cache misses are shown in \figurename~\ref{dal-ration}. For most of the benign programs, MSHR misses account for less than $3.75\%$ among all cache accesses, except for the {\tt mcf\_s} program accounting for about $10.83\%$. However, the proportion of MSHR misses among all cache accesses exceeds $20\%$ for the Spectre-type attack programs. Similarly, for all benign programs, the proportions of MSHR misses among all cache misses are less than $60\%$. However, the proportions of MSHR misses among all cache misses are much higher for all the attack programs (more than $96\%$). Overall, due to the frequent requirement for cache eviction and observation, cache side-channel attack programs tend to have more MSHR misses than benign programs.

 {\textbf{Observation 2: Two limitations of conflict-based attackers.} Firstly, the attacker is unable to distinguish whether the cache state change is due to victim activities or hardware prefetching. If a hardware prefetching occurs during an attack, the attacker can only distinguish between a cache hit or miss; however, it has no way of determining the cause of this change. Secondly, the attacker cannot perceive multiple operations on the same cache line. Therefore, it is possible to hide or distort the victim's exact cache access pattern through additional operations, such as modifying the cache line replacement priority or bringing back evicted cache lines.}


\section{Threat Model}\label{section:Threat}

The goal of PCG is to utilize prefetching to mitigate conflict-based cache side-channel attacks, offering a low-cost and efficient alternative to complex secure cache design solutions. Moreover, we exclusively focus on the scenarios where both the attacker and the victim execute instructions in the same core. Cross-core conflict-based cache side-channel attacks can be mitigated using the BITP~\cite{BITP2019}, which offers a low-cost and efficient cross-core prefetching-based solution. We assume that the attacker has the following capabilities:
\begin{itemize}
    \item The knowledge of cache indexing and replacement policies. So the attacker can install or evict any cache lines and incur MSHR misses during its attacking process.
    \item The capability of constructing conflict-based cache side channels based on the state of cache occupancy.
    \item The ability to obtain response latency for memory access operations. It reveals whether data appears in the cache hierarchies, possibly as a result of the victim's access.

\end{itemize}

\mypara{Out of scope} Although other hardware units such as Instruction Cache~\cite{ICacheAttack}, TLB~\cite{TLB2018}, MSHR~\cite{Interference2021}, execution units~\cite{SMoTherSpectre2019}, and branch predictors~\cite{NDA2019,BranchScope2018} can also be utilized to construct microarchitectural side channels, we confine our protection scope solely to data caches. This decision is based on the fact that the data cache is one of the hardware units that are most frequently utilized to construct side channels due to its persistent state and high bandwidth~\cite{hu2023protecting}.  {Additionally, simultaneous multi-threading is beyond the scope of this paper.} 



\section{PCG Design}\label{section:Design}
This section provides the design overview of PCG and the design details of its two components, Attack Aware Module (AAM) and Observation Confused Module (OCM).

\subsection{Overview}
\label{subsection:overview}

\begin{figure}[!t]
  \centering
  \includegraphics[width=0.9\linewidth]{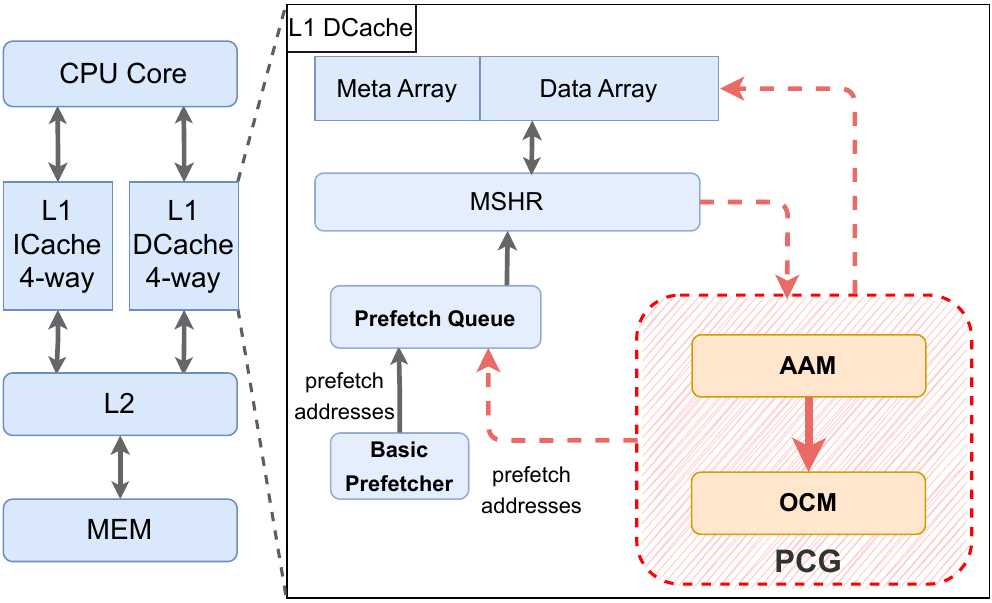}
  \caption{Overview of PCG design.}
  \label{overview}
\end{figure}

PCG focuses on mitigating conflict-based cache side-channel attacks targeting the data cache, which provides a primary attack surface~\cite{RandomFillCache2014} and main side channel utilized in existing speculative execution attacks~\cite{XiongSurvey2021}. To simplify the explanation without loss of generality, we take the PCG implemented in the L1 DCache for example. The implementation in other-level cache (e.g., L2 Cache) is similar. As shown in \figurename~\ref{overview}, the red dashed square and arrows represent PCG and its connections with other components in L1 DCache, respectively.


AAM in PCG utilizes MSHR misses to identify the abnormal cache sets that may be exploited by attackers. Moreover, with the AAM's identification (corresponding to the \textit{dangerSet} register set by AAM), OCM utilizes data prefetching to reduce the victim-related cache footprints and to add victim-unrelated cache accesses as interfering noise, based on which the attacker's observation of cache state can be efficiently confused.

Note that PCG can be readily integrated with the basic prefetchers without changing any of their original input or output signal ports, simply by appending its prefetching requests to the Prefetch Queue\footnote{The Prefetch Queue is a first-in, first-out structure used to store prefetch addresses and to send the memory request that contains the prefetching address to the cache. If the prefetch address is already present in the cache, it is ignored; otherwise, the memory request is sent to the next-level cache.}, which is shared with the basic prefetchers. The primary function of a basic prefetcher is to improve performance, while PCG is used to enhance security.

\subsection{{AAM Design}}

The AAM's configuration is related to the cache. Let $S$ and $W$ denote the number of cache sets and ways of L1 DCache, respectively. The structure of AAM is shown on the right in \figurename~\ref{aam}. AAM only handles the cache accesses that result in MSHR misses in order to more accurately capture attacker behavior. An example of valid cache accesses is depicted on the left in \figurename~\ref{aam}. For each cache access, AAM has a corresponding pair of \textit{PC} and \textit{Addr} as its one input. We introduce \textit{accessCounter} with $S$ entries to record the number of access times of each cache set.  The \textit{lastPC} register is used to record the \textit{PC} in the previous input. The $S$-bit \textit{dangerSet} register is used to identify abnormal cache sets, which is initialized to zero. Specifically, the working principles of AAM are as follows:

\begin{figure}[!t]
  \centering
  \includegraphics[width=\linewidth]{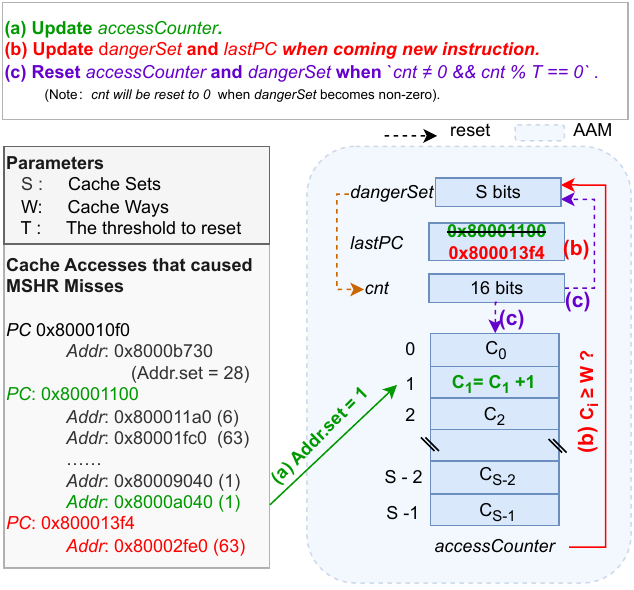}
  \caption{The design of AAM.}
  \label{aam}
\end{figure}

\mypara{(a) Update \textit{accessCounter}}
With the \textit{Addr} in an input, the index of the accessed cache set can be obtained, denoted as $i$. Then the $i^\mathrm{th}$ entry of \textit{accessCounter}, denoted as $C_i$, will be updated as follows:
\begin{equation}
\label{equ5}
    C_i = (C_i == W)\ ?\ W : C_i + 1,\quad C_i\in [0,W]~.
\end{equation}%

Taking $\mathit{Addr=0x8000a040}$ shown in \figurename~\ref{aam} for example, the cache set index corresponding to the $6^{th}$--$11^{th}$ bits of the address is $1$. Therefore, $C_1$ will be updated accordingly.

\begin{algorithm}[!t]
\caption{OCM prefetching algorithm}\label{alg-1}
\LinesNumbered
\SetAlgoNoLine
\KwIn{\textit{coreReqAddr}, \textit{refSet}}
\KwOut{\textit{prefetchQueue}}
$\textbf{E}\leftarrow$  the cache entry allocated to \textit{coreReqAddr}\;
$i \leftarrow coreReqAddr.set$\;\label{alg1:currAddrSet}

\tcc {\textbf{Reduce victim-relevant cache footprints}}
\If{$dangerSet[i] == 1$ {\rm and a cache line is evicted }}{\label{alg1:checkDan}
    set $\textbf{E}$ with the highest replacement priority\;\label{alg1:setVictAcc}
    \textit{eviAddr} $\leftarrow$ the address of the evicted cache line\;\label{alg1:eviAddr}
    $prefetchQueue.push\_back(eviAddr)$\;\label{alg1:eviAddrPref}
}\label{alg:less2}

\tcc {\textbf{Add victim-irrelevant cache footprints}}
$refSet[i] \leftarrow 1 $\;\label{alg1:updateRefSet}
\If{Cache Miss}{\label{alg:more1}
       $blkAddr \leftarrow blockAddress(coreReqAddr)$\;
       $tempAddr \leftarrow coreReqAddr$\;
    \For{$d = 1$ \KwTo $degree$}{
        $direction \leftarrow $ Randomly select from $ \left\{0, 1 \right\}$\;\label{alg1:direction}
        \uIf{direction == 0}{\tcc {forward prefetching}
            $tempAddr \leftarrow  blkAddr + d \times BlockSize$\;
        }\Else{\tcc {backward prefetching}
             $tempAddr \leftarrow  blkAddr - d \times BlockSize$\;
        }\label{alg1:GenPrefAddrEnd}
        \tcc{The function below is detailed in Algorithm~\ref{alg-2}}
        $prefAddr\leftarrow{}BalancedSet(tempAddr,refSet)$\;
        $prefetchQueue.push\_back(prefAddr)$\;
    }
}\label{alg:more2}

\label{alg1}
\end{algorithm}

\mypara{(b) Update \textit{dangerSet}} When \textit{PC} differs from \textit{lastPC}, i.e., a new instruction is encountered, \textit{dangerSet} will be updated. Each bit $D_i$ (denoted as the $i^\mathrm{th}$ bit of \textit{dangerSet}) is updated as:
\begin{equation}
\label{equ7}
    D_i = D_i\ ||\ (C_i \geq \tau),\ D_i\in \{0,1\}~,
\end{equation}
where $\tau$ is the counting threshold of $C_i$. In this paper, we set $\tau=W$, since the attacker has to access the cache at least $W$ times to occupy a target set. If $D_i=1$, it indicates that ${\rm s}_i$ is an abnormal cache set. 

\mypara{(c) Reset \textit{accessCounter} and \textit{dangerSet}} \textit{accessCounter} and \textit{dangerSet} need to be refreshed at a certain period, denoted as $T$. To this end, a 16-bit counter, \textit{cnt}, is employed, which increases by one at every clock tick. When $cnt\neq0$ and $cnt\ \%\ T == 0$, AAM will reset \textit{accessCounter} and \textit{dangerSet} to zero. It is worth noting that \textit{cnt} is reset when \textit{dangerSet} becomes a non-zero value to prevent \textit{dangerSet} from being cleared before it can be used by the OCM.

The selection of $T$ is essentially based on the trade-off between performance and security. If $T$ is set too large, as the clock cycles increase, some benign cache sets may be misjudged as abnormal. If the threshold is set too small, some abnormal cache sets may be missed. We will discuss the impact of different reset periods on performance and security in Section \ref{subsection-sensi}.

\subsection{OCM Design}
OCM reduces victim-relevant cache footprints and adds victim-irrelevant cache footprints as interfering noise, to confuse the attacker's observation. The algorithm is summarized in Algorithm~\ref{alg-1}.

\mypara{(a) Reduce victim-relevant cache footprints}
 {In this paper, we define the victim-relevant cache footprints as the cache occupancy changes caused by the victim's installation and eviction of cache lines. When the request address from the core (denoted as $coreReqAddr$) is 1) mapped to an abnormal cache set and 2) causes a cache line to be evicted, OCM considers it a potential victim request (Line~\ref{alg1:checkDan} Algorithm~\ref{alg-1}).  We then reduce the victim's footprints by combining the following two ways:}
\begin{itemize}
    \item Set \textbf{E}, the cache entry assigned to the potential victim request, with the highest replacement priority (Line~\ref{alg1:setVictAcc} Algorithm~\ref{alg-1}). Thus, the victim line will be evicted from the cache as soon as possible.
    \item Push \textit{eviAddr}, the address of the evicted cache line, into the Prefetch Queue (Lines~\ref{alg1:eviAddr}--\ref{alg1:eviAddrPref} Algorithm~\ref{alg-1}). OCM will bring back the line evicted by the victim's activity back to the cache.	
\end{itemize}

Overall, those lines possibly evicted due to the victim's activity will be brought back into the cache through prefetching, and those lines installed into the cache by the victim will be evicted from the cache as early as possible. Both of these actions reduce the victim's footprints residing in the cache.

Due to the latency of prefetching and the uncertainty of when the attacker will access the cache, some footprints of the victim may still remain in the cache. To protect the remaining victim footprints from being extracted by the attacker, PCG employs prefetching to introduce victim-irrelevant cache footprints as interference noise. Thus, it is difficult for the attacker to extract the victim's activities from the obfuscated cache access patterns (see the experimental results shown in Section~\ref{subsection:breaking}).

\begin{algorithm}[!t]
\caption{$BalancedSet(\cdot)$}\label{alg-2}
\LinesNumbered
\SetAlgoNoLine
\KwIn{$tempAddr$, \textit{refSet}}
\KwOut{$prefetchAddr$}
    $prefetchAddr \leftarrow tempAddr$\;
    \If{$refSet == 11\ldots1$}{\label{alg2:reset}
        $refSet \leftarrow 00\ldots0$\;\label{alg2:iniRef}
        $danSet \leftarrow \sim dangerSet$\; \label{alg2:iniDan}
    }
    $t\_set \leftarrow tempAddr.set$\;
    \uIf{$danSet \neq 11\ldots1$}{
        $final\_set \leftarrow $ closest $s$ to $t\_set$ with $danSet[s] = 0$\;\label{alg2:danSet}
        $danSet[final\_set] \leftarrow 1$\;
    }\Else{
        $final\_set \leftarrow $ closest $s$ to $t\_set$ with $refSet[s] = 0$\;\label{alg2:refSet}
        $refSet[final\_set] \leftarrow 1$\;
    }
    $prefetchAddr.set \leftarrow final\_set$

\label{alg2}
\end{algorithm}

\mypara{(b) Add victim-irrelevant cache footprints}
OCM triggers prefetching and adjusts the cache set that the prefetching address is mapped to, introducing balanced, noisy victim-irrelevant cache footprints.

When a cache miss occurs, the forward or backward memory addresses adjacent to $coreReqAddr$ are generated (Lines~\ref{alg1:direction}--\ref{alg1:GenPrefAddrEnd} Algorithm~\ref{alg-1}), and the number of these addresses is determined based on the prefetching degree, denoted as degree. 
The generated address, referred to as \textit{tempAddr}, is adjusted to the closest address that is mapped to an unaccessed cache set, referred to as \textit{final\_set} before being pushed into the Prefetch Queue, as shown in Algorithm~\ref{alg-2}.

We use two $S$-bit registers, \textit{danSet} and \textit{refSet}, to record the currently accessed abnormal cache sets and the currently accessed cache sets, respectively. The result of the bitwise inverse of the \textit{dangerSet} is used to initialize the \textit{danSet} (Line~\ref{alg2:iniDan} Algorithm~\ref{alg-2}). \textit{refSet} is initialized to zero (Line~\ref{alg2:iniRef} Algorithm~\ref{alg-2}). OCM updates the \textit{refSet} based on the cache set mapped by $coreReqAddr$ (Lines~\ref{alg1:currAddrSet} and \ref{alg1:updateRefSet} Algorithm~\ref{alg-1}). When all cache sets are marked as accessed by \textit{refSet} (Line~\ref{alg2:reset}  Algorithm~\ref{alg-2}), \textit{danSet} and \textit{refSet} are reset. The cache set mapped by \textit{tempAddr} is denoted as \textit{t\_set}. First, OCM tries to set \textit{final\_set} to the closest unaccessed abnormal cache set to \textit{t\_set} (Line~\ref{alg2:danSet} Algorithm~\ref{alg-2}). If all abnormal cache sets have been marked as accessed by danSet, OCM sets \textit{final\_set} to the closest unaccessed cache set to \textit{t\_set} (Line~\ref{alg2:refSet} Algorithm~\ref{alg-2}).

The prefetching degree should be carefully tuned. On one hand, the degree should be large enough to ensure the generation of enough noise, making it difficult for the attacker to pinpoint the exact cache set accessed by the victim. On the other hand, too aggressive prefetching may introduce more useless cache lines which pollute the cache. Through experimentation, we found that a degree of 4 is a good choice, which allows for maintaining performance while making the cache behavior more unpredictable.

\section{Security Analysis}

In this section, we discuss the security of PCG and compare it with the previous prefetching-based defenses, including PREFENDER and DP. We briefly describe how PCG defends against enhanced attackers which can suppress the prefetcher.
\label{section:Security}
\subsection{Conflict-based Cache Side-channel Attacks}

\textbf{Case 1: Attackers do not know PCG design.} For the known conflict-based cache side-channel attacks such as Evict+Reload, Evict+Time, and Prime+Probe, attackers frequently access ES during Phase 1 to occupy the target cache set, resulting in numerous MSHR misses. Consequently, AAM is capable of identifying the target cache set as anomalous.

For Evict+Reload, the key to success lies in the attacker's ability to observe the cache hit due to the victim accessing the target cache block. However, the victim cannot determine whether the target address hit is due to the victim accessing or due to prefetching. Meanwhile, the victim is also unable to determine whether the cache miss is due to the victim not accessing the corresponding cache block or due to the victim's line being evicted before it reloads.

For Evict+Time, if the attacker observes a longer execution time of the victim when target cache blocks have been evicted, the attack succeeds. However, the attacker (which causes numerous cache misses) triggers OCM prefetching, which will compensate for the execution delay caused by the eviction of the target block~\cite{BITP2019}. Thus, the execution time of the victim does not change much, leading to the failure of the attack.

For Prime+Probe, the attacker will be successful if it observes a cache miss when probing its own lines, indicating the victim has evicted the corresponding cache line. However, when prefetching from the OCM leads to the eviction of an attacker's line, it also results in the attacker experiencing a cache miss. Furthermore, OCM may bring back the cache line evicted due to victim activity. This means that even if the victim accesses the target address, the attacker observes a cache hit instead of a cache miss.

\textbf{Cache 2: Attackers know PCG design in advance.}
We must admit that attackers may know PCG design in advance and perform an adaptive attack to bypass AAM. For instance, attackers could accurately execute Phase 1 within two reset periods, preventing the target cache set from being identified as abnormal. However, we believe this not only increases the cost of the attack but also makes it challenging to implement in real systems. Even if PCG fails to capture abnormal cache sets in such scenarios, OCM still works all the time, triggering prefetches throughout the cache. In other words, even in the worst case, the security of the PCG is comparable to that of the DP.

\subsection{Security Comparison with Recent Works}
\label{subsection:securityComparison}
\begin{figure*}[!t]
  \centering
  
  \subfloat[Baseline (1 Round)]{\includegraphics[width=0.24\textwidth]{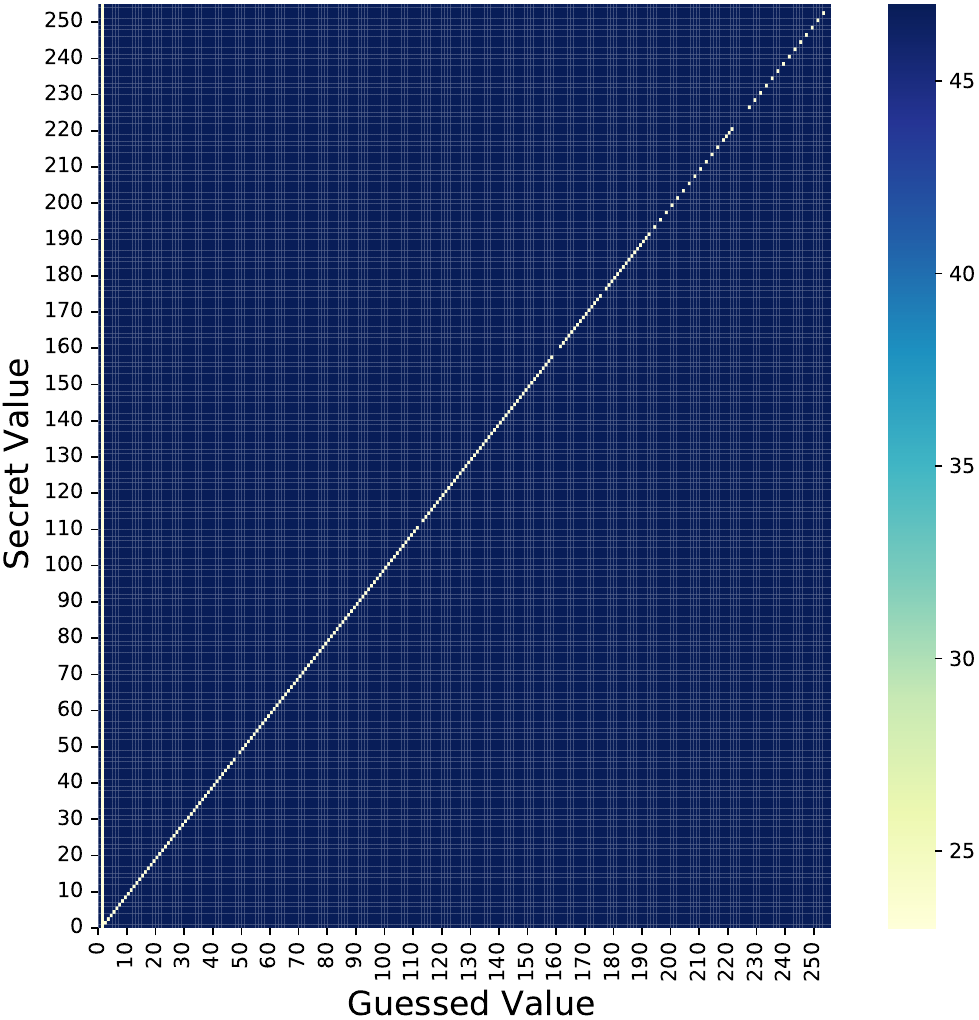}\label{subfig:hotmap-base}}
  \hfill 
  \subfloat[PREFENDER (1 Round)]{\includegraphics[width=0.24\textwidth]{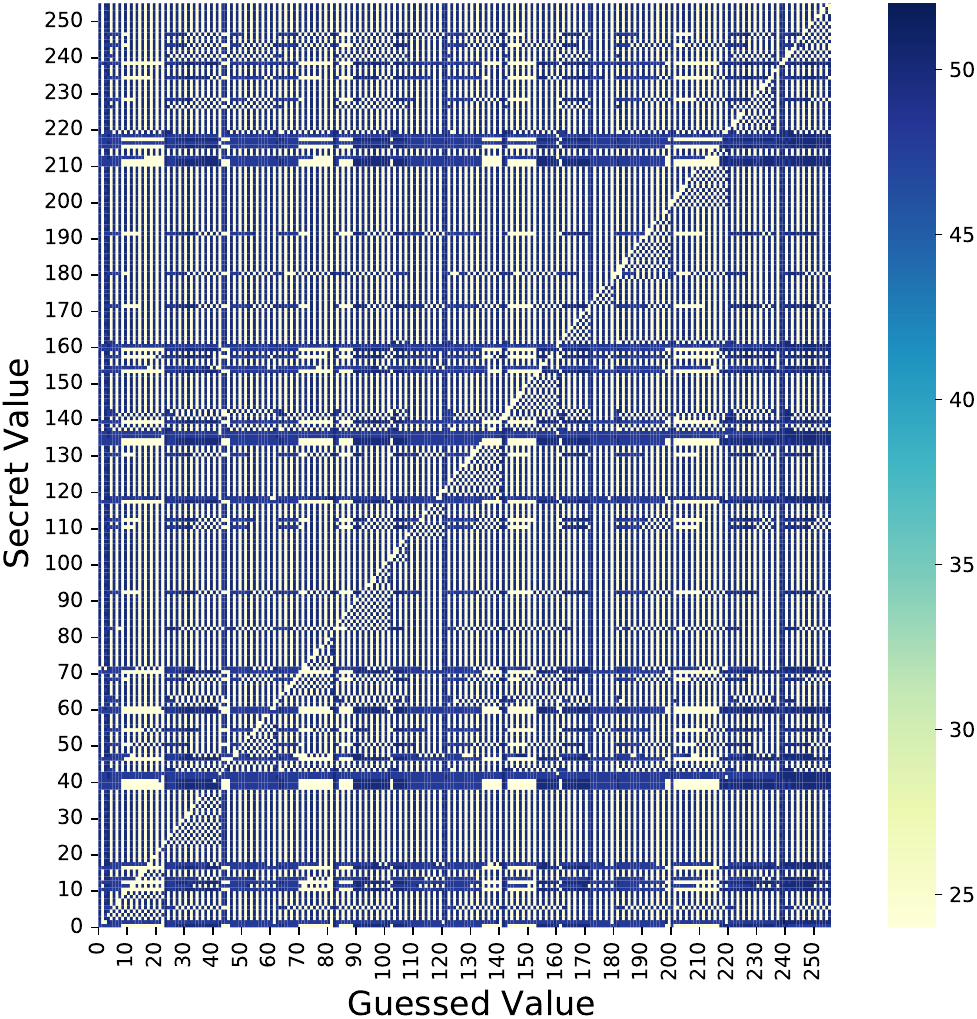}\label{subfig:hotmap-pref}}
  \hfill 
\subfloat[DP (1 Round)]{\includegraphics[width=0.24\textwidth]{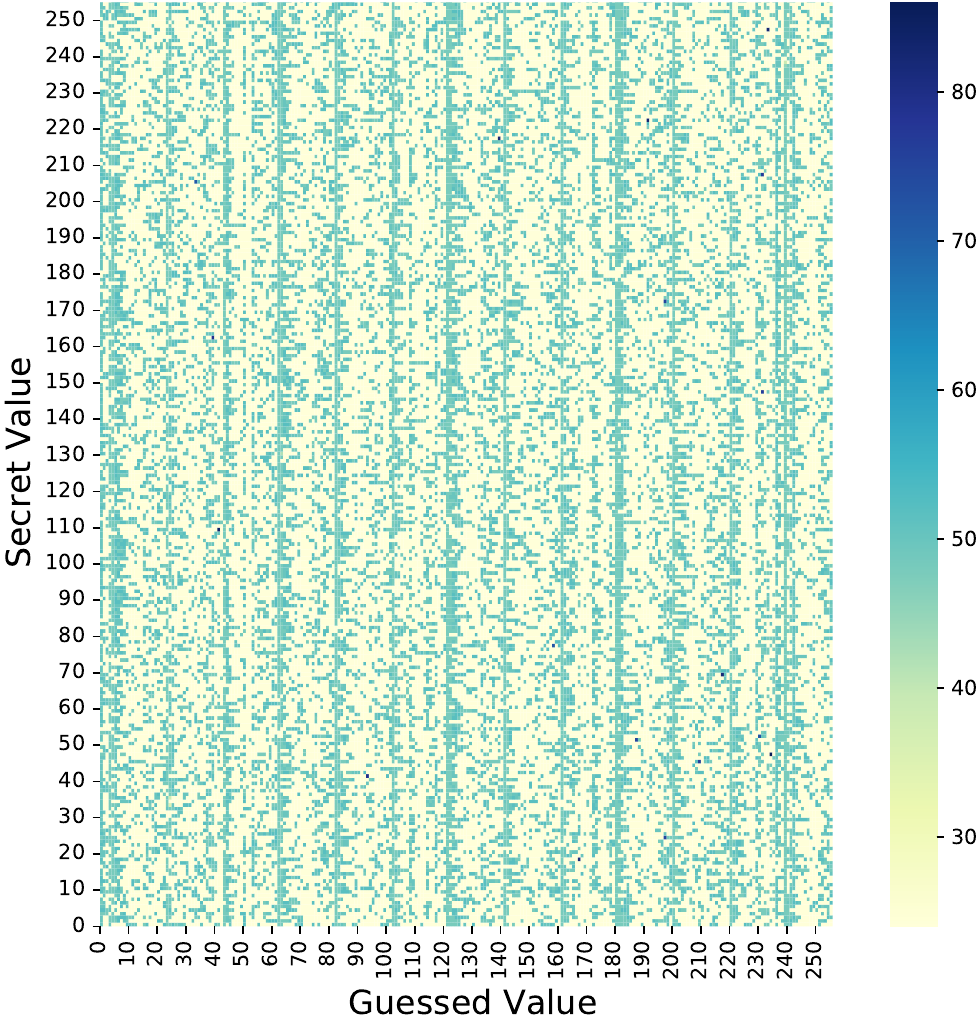}\label{subfig:hotmap-dis}}
  \hfill 
  \subfloat[PCG (1 Round)]{\includegraphics[width=0.24\textwidth]{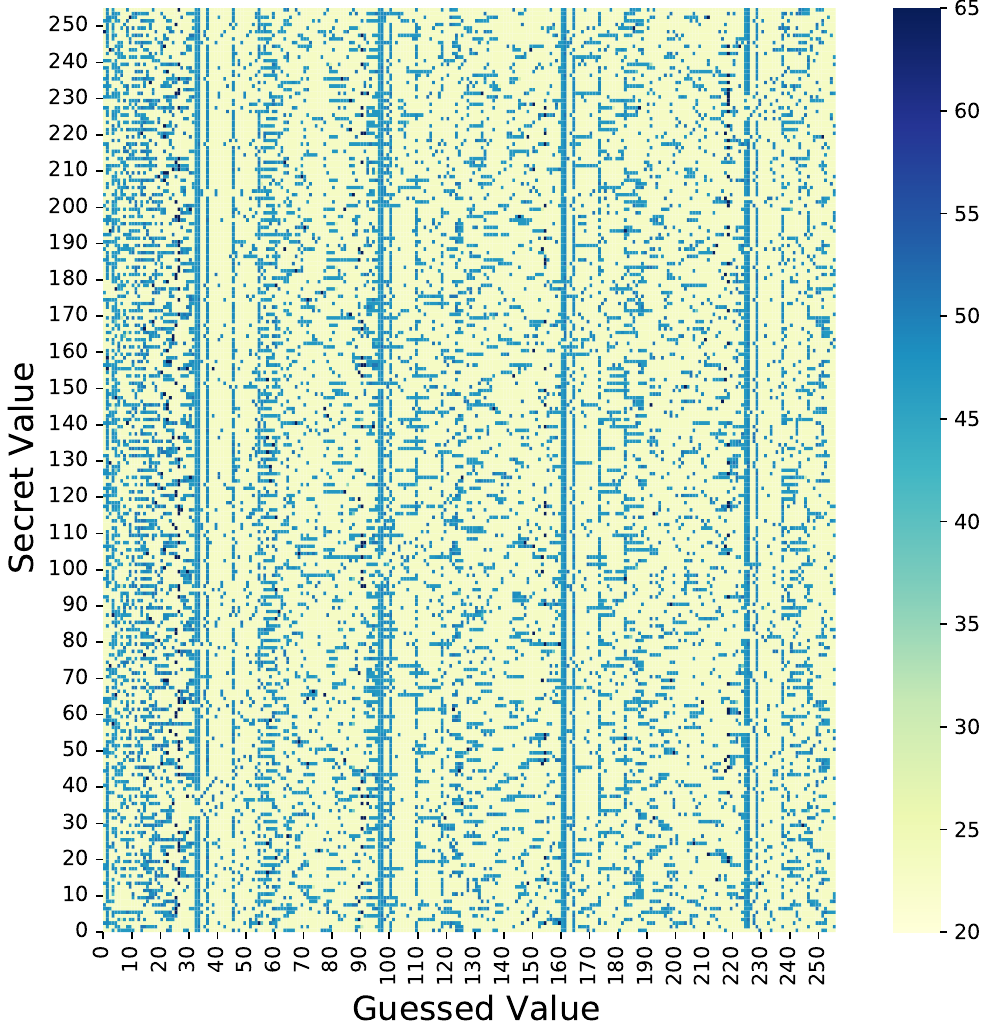}\label{subfig:ht-pcg-1}}
  
    \subfloat[PREFENDER (100 Rounds) ]{\includegraphics[width=0.24\textwidth]{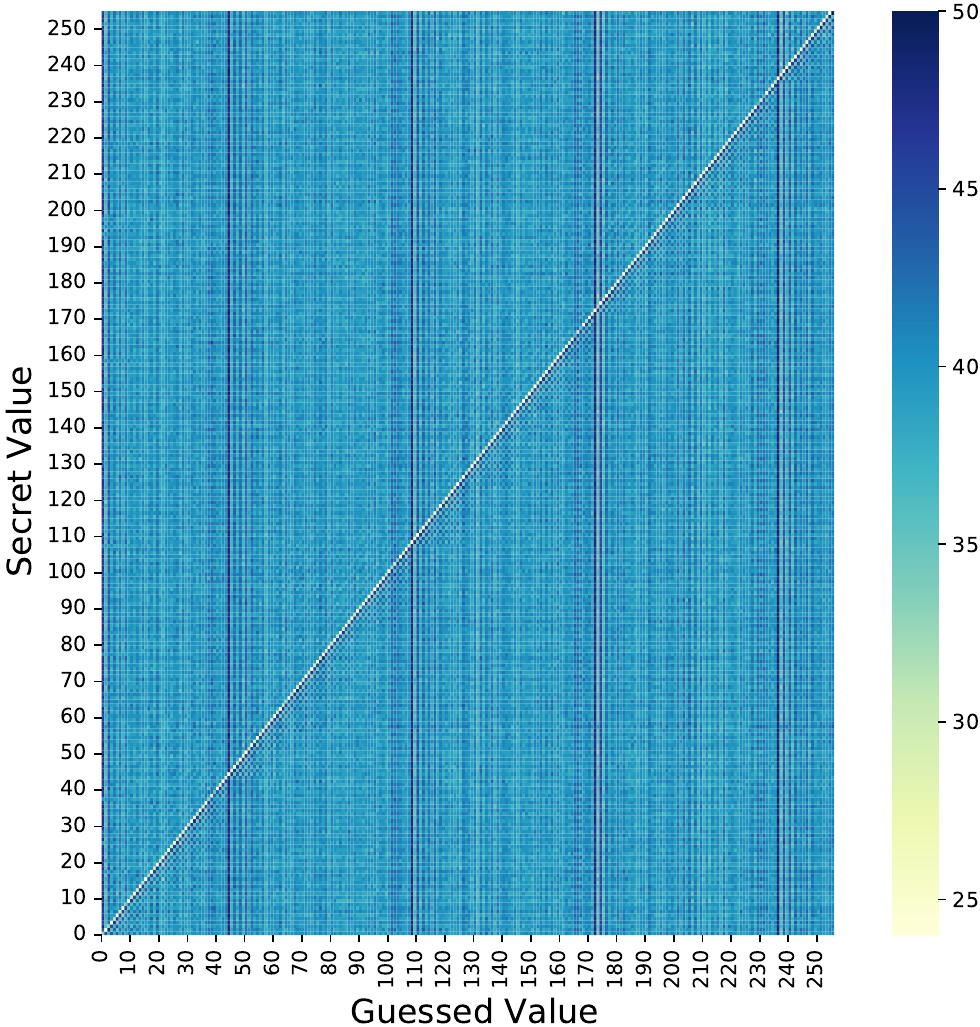}\label{subfig:ht-pref-100}}
  \hfill 
  \subfloat[DP (100 Rounds)]{\includegraphics[width=0.24\textwidth]{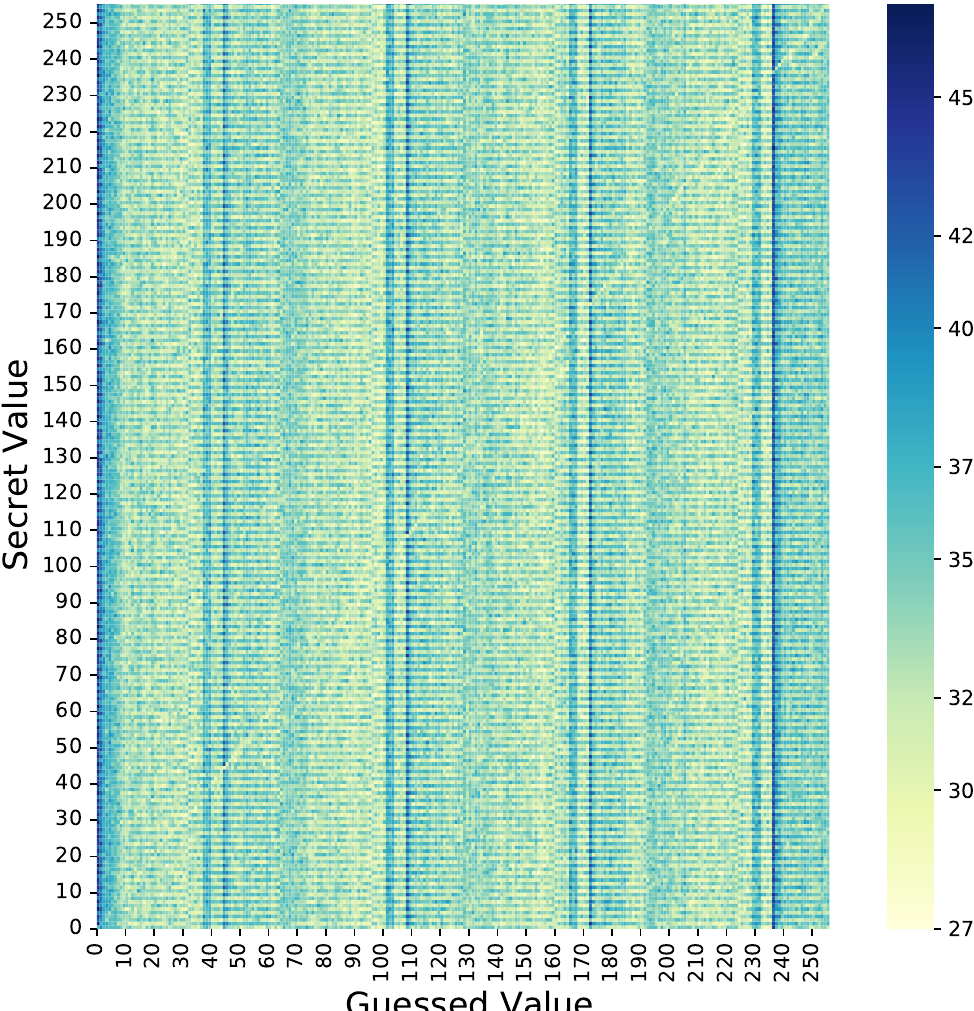}\label{subfig:ht-dis-100}}
  \hfill 
\subfloat[PCG (100 Rounds)]{\includegraphics[width=0.24\textwidth]{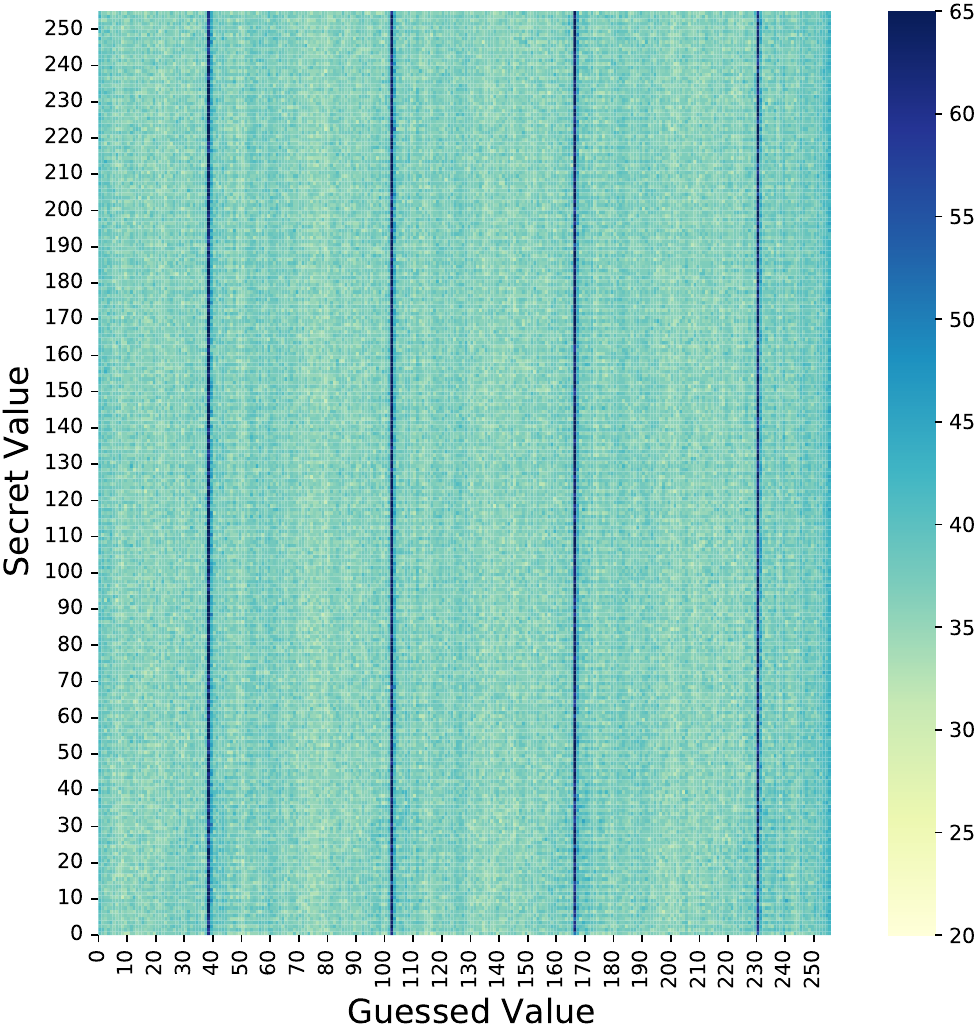}\label{subfig:ht-pcg-100}}
  \hfill 
  \subfloat[PCG (1,000 Rounds)]{\includegraphics[width=0.24\textwidth]{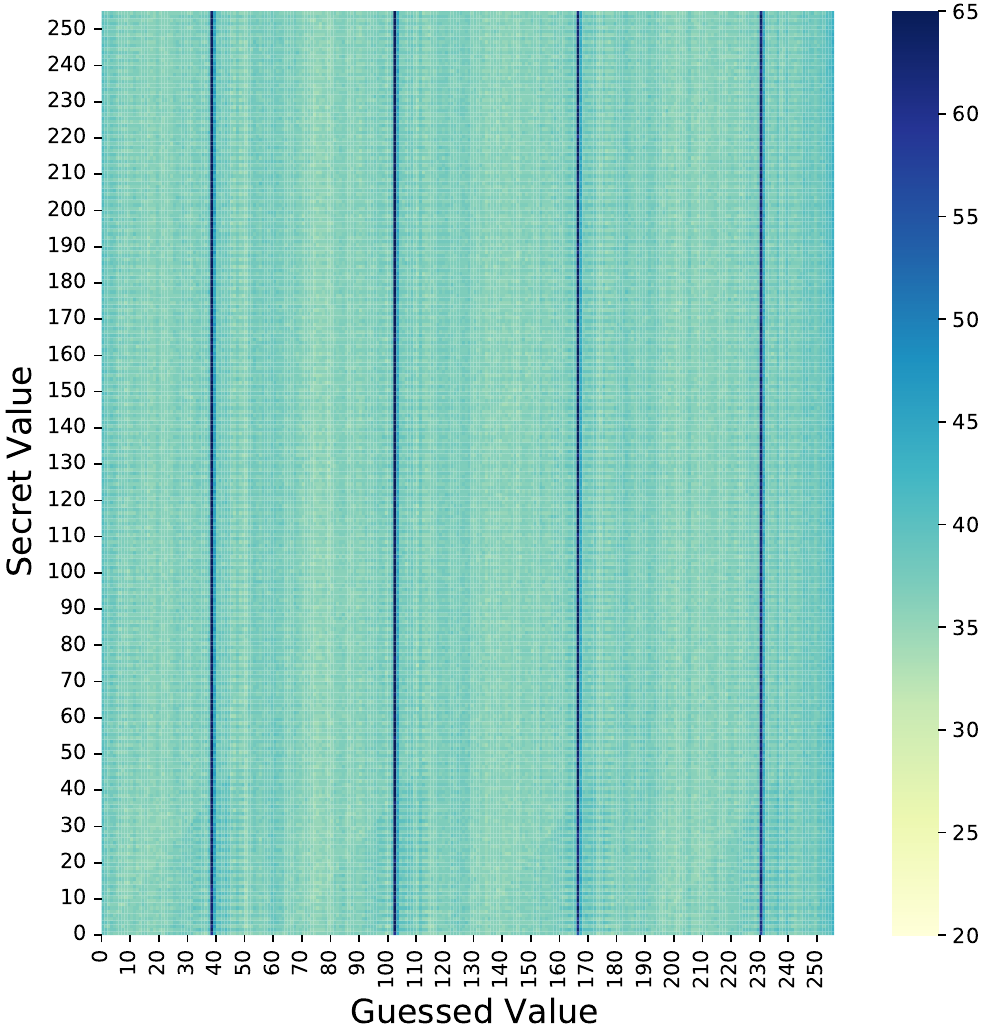}\label{subfig:ht-pcg-1000}}
   
  
  \caption{Security comparison of previous works with PCG.} 
  \label{pic.hotmap}
\end{figure*}

Previous prefetching-based defense mechanisms, including PREFENDER and DP, simply utilize prefetching to add noise to the cache without reducing victim's cache footprints. When attackers use different cache probing sequences, it will result in PREFENDER and DP prefetching different addresses. This leads to the fact that an attacker can reliably distinguish between noise and real victim cache lines with enough attacks. Instead, PCG not only adds victim-irrelevant cache footprints but also reduces victim-relevant cache footprints. Thus providing a higher level of security than previous work.

\lstset{escapeinside={(*@}{@*)}}
\begin{lstlisting}[language=C, caption=A PoC code of the Evict+Reload attack., label=list1, upquote=true]
uint8_t secretValue;
uint8_t array2[256 * BlockSize];//side-channel(SC) (*@\label{list:array2size}@*)
uint64_t accessCycles[256];(*@\label{list1:auxiliary}@*)

//Phase 1: Initializing the side-channel
EvictFromCache(array2); (*@\label{list:clearcache}@*) 

//Phase 2: Access and Send Secret Data
secretValue was accessed during victim exectution;(*@\label{list1:accessSecret}@*)
uint8_t dummy = array2[secretValue*BlockSize];(*@\label{list1:accessArr2}@*)

//Phase 3: Recovering the secret
for (uint8_t guess  = 0 to 255)(*@\label{list1:Phase3}@*) {(*@\label{list1:forInPhase3}@*)
    //Obtain the current clock cycle:
    uint64_t start = rdcycle();
    uint8_t dummy = array2[guess*BlockSize];
    accessCycles[guess] = rdcycle() - start;(*@\label{list:traverse}@*)
}(*@\label{list1:Phase3End}@*)
//Find the index of the minimum value in accessCycles, which is the secret value
uint8_t secretVal = GetIdxMinVal(accessCycles);(*@\label{list1:secret}@*)

\end{lstlisting}

We further conduct experiments in gem5 to illustrate the defensive capabilities of PREFENDER, DP and PCG against conflict-based cache side-channel attacks. Specifically, we execute an intra-domain Evict+Reload attack on L1 DCache. The L1 DCache configuration in gem5 is shown in Table~\ref{tab-gem5}. Listing \ref{list1} depicts the Proof of Concept (PoC) code for the Evict+Reload attack employed in this paper. This code simulates the sequential execution of attacker and victim, allowing the attacker to carry out the attack arbitrarily, which further advantages the attacker.

During Phase 1, the attacker evicts \textit{array2} from the cache by constructing set conflicts. In Phase 2, the victim loads the \textit{secretValue} (Line~\ref{list1:accessSecret} Listing~\ref{list1}) and accesses a specific position within \textit{array2} shared with the attacker (Line~\ref{list1:accessArr2} Listing~\ref{list1}), which depends on the \textit{secretValue}. In Phase 3, the attacker makes guesses about all possible values (8 bits data, 256 possibilities) by sampling the access latency at the corresponding position in \textit{array2} (Lines~\ref{list1:Phase3}--\ref{list1:Phase3End} Listing~\ref{list1}). Among all the entries in \textit{array2}, the one which has the shortest access latency is associated with the secret value (Line~\ref{list1:secret} Listing~\ref{list1}). 

Each round experiment contains 256 attacks on different secret values. \figurename~\ref{pic.hotmap} shows the results of 1-round (\figurename~\ref{subfig:hotmap-base}--\ref{subfig:denoise-pcg}), 100-rounds (\figurename~\ref{subfig:ht-pref-100}--\ref{subfig:ht-pcg-100}), and 1,000-rounds (\figurename~\ref{subfig:ht-pcg-1000}) experiments, resulting in heat maps that give the average access latency for each pair of $\langle$secret value, guessed value$\rangle$. Brighter points (in yellow) represent shorter access latency. The more obvious the diagonal shows, the worse security the corresponding scheme has, as it indicates that the attacker is more likely to infer the secret value from the access latency of \textit{array2}.

For the baseline, \figurename~\ref{subfig:hotmap-base} shows a clear diagonal that links the secret value to the index of \textit{array2} (the guessed value). In other words, the attacker can infer the secret value by determining which location in \textit{array2} has the shorter access latency. As shown in \figurename~\ref{subfig:hotmap-pref} and \ref{subfig:ht-pref-100}, despite PREFENDER causing cache hits at multiple index positions in \textit{array2}, both 1-round and 100-rounds experiments display clear diagonal patterns. DP exhibits a noisy heat map in 1-round experiment (\figurename~\ref{subfig:hotmap-dis}), but when the number of experiments is increased to 100 rounds (\figurename~\ref{subfig:ht-dis-100}), a faint diagonal pattern can still be discerned. This may reveal the actual cache access pattern. When PCG is enabled, even after 1,000-rounds experiments (with the number of attacks increased to 256,000), which takes about 13 hours, the attacker is still unable to extract any meaningful information, as shown in \figurename~\ref{subfig:ht-pcg-1000}.

The heat maps of PREFENDER and DP suggest that the noise introduced by prefetching alone is insufficient to defeat conflict-based side-channel attacks, which can sample cache state multiple times. Instead, PCG introduces noise while reducing the victim's cache footprints, providing stronger security compared to PREFENDER and DP. 

\subsection{\textbf{Breaking PREFENDER and DP}}
\label{subsection:breaking}

\begin{figure*}[!t]
  \centering
  \subfloat[PREFENDER]{\includegraphics[width=0.33\textwidth]{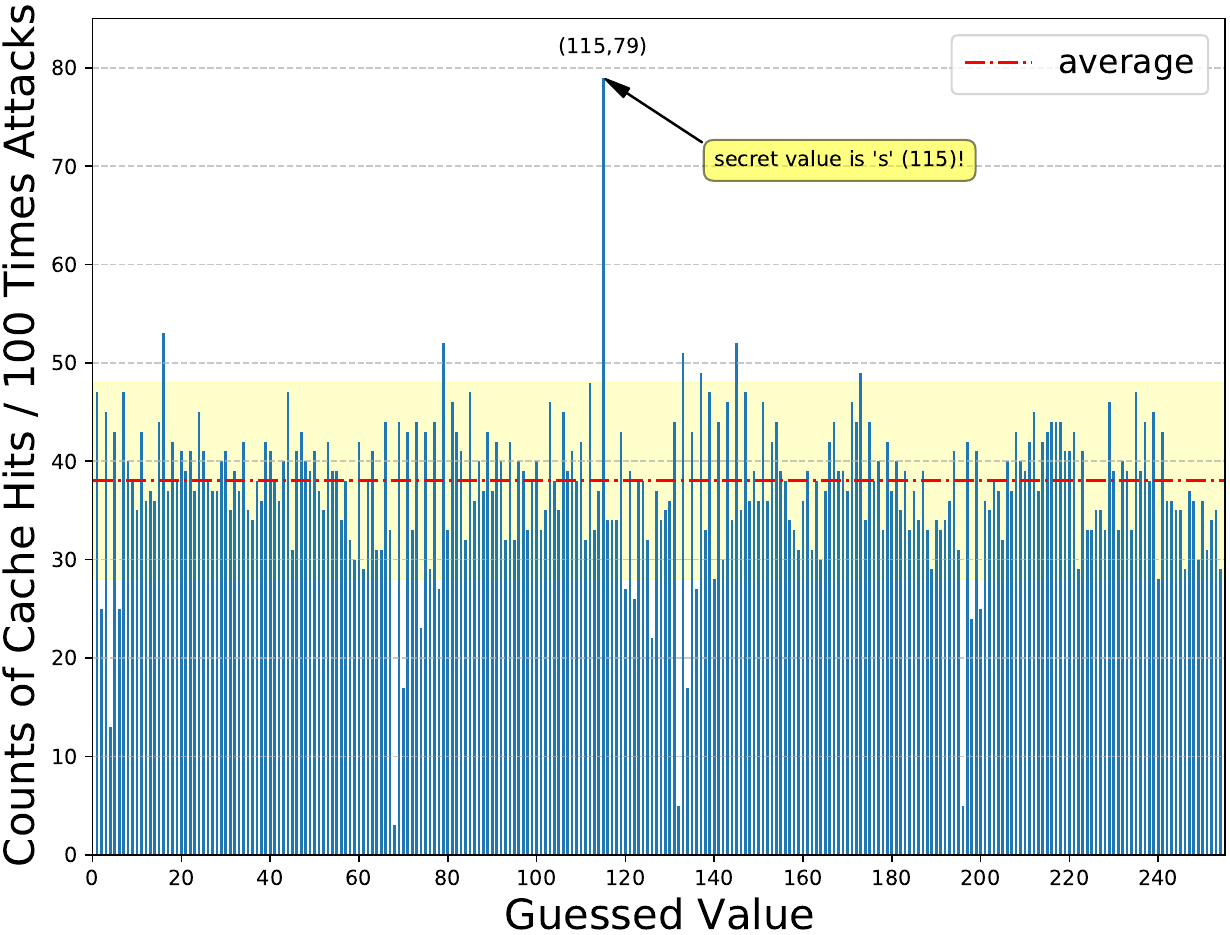}\label{subfig:denoise-pref}}
  \hfill
  \subfloat[DP]{\includegraphics[width=0.33\textwidth]{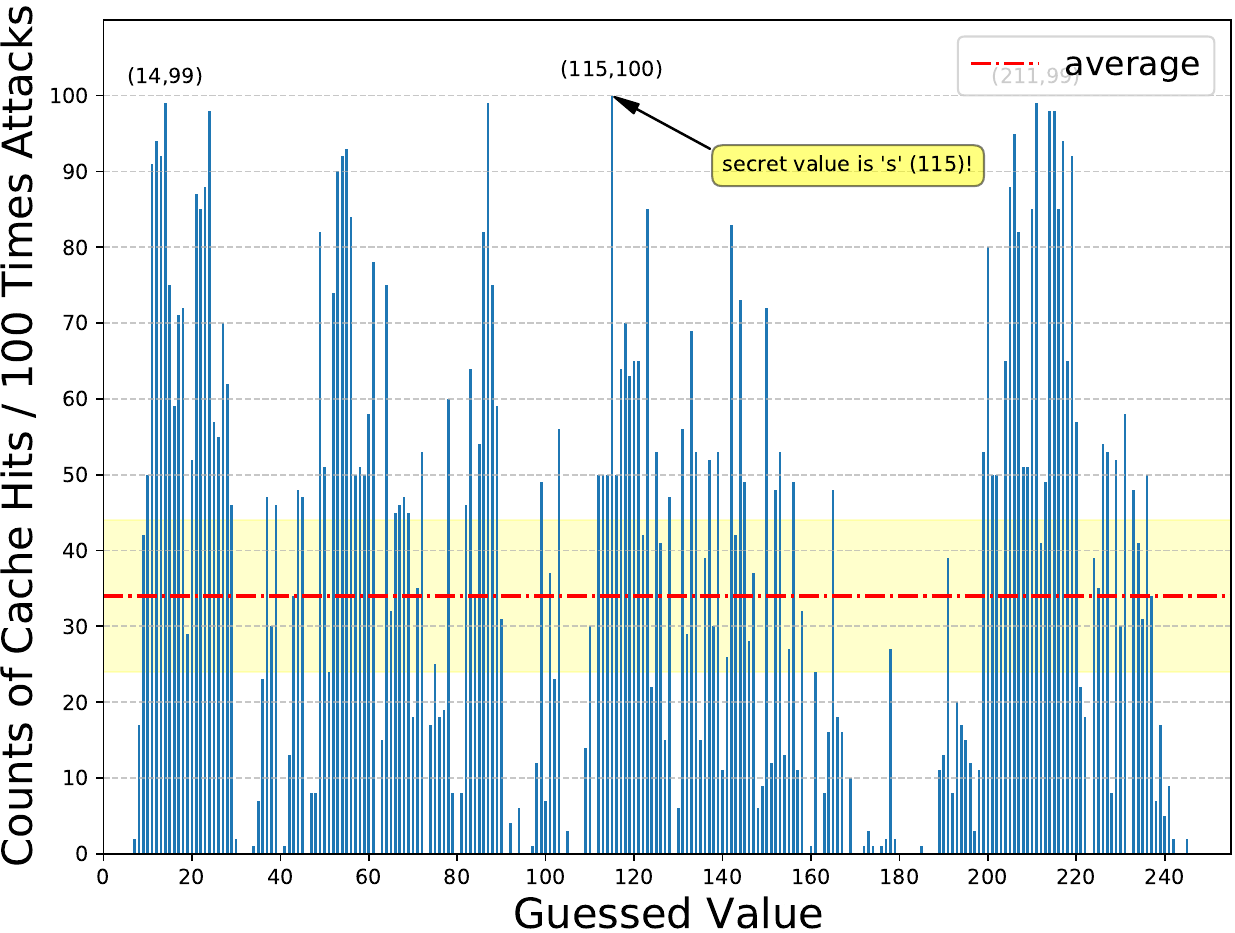}\label{subfig:denoise-dis}}
  \hfill
\subfloat[PCG]{\includegraphics[width=0.33\textwidth]{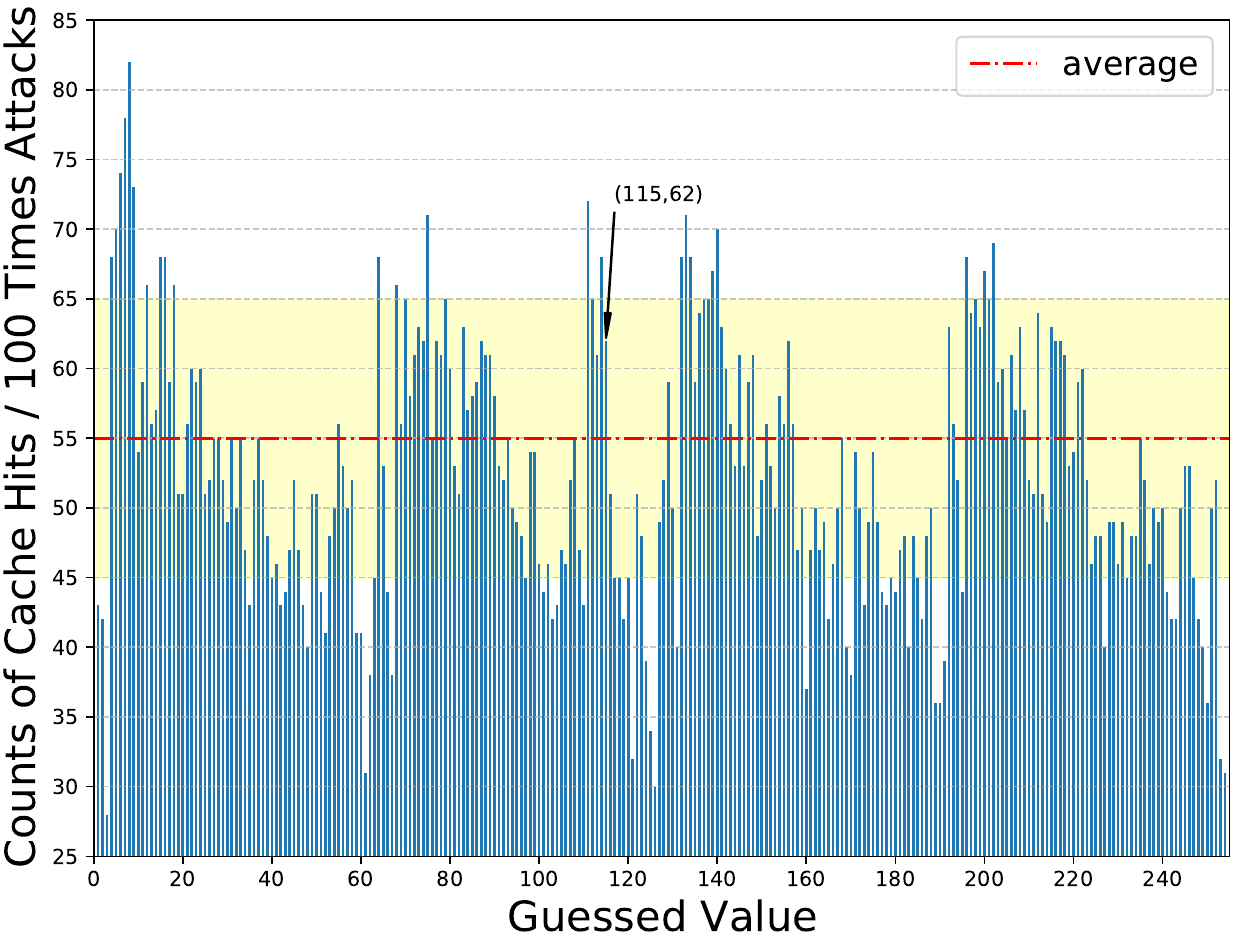}\label{subfig:denoise-pcg}}
  
  \caption{Repeat the attack 10,000 times to recover the secret `s'. Neither PREFENDER nor DP can withstand the repeated attacks, and `s' is determined with the highest cache hits.}
  \label{pic.denoise}
\end{figure*}

We demonstrate that PREFENDER and DP can be broken through a sufficient number of attacks. We conduct 10,000 attacks in gem5 to obtain the secret character `s', and every 100 attacks the probe sequence of \textit{array2} will be changed. This is to generate different memory access histories, which can lead to different addresses to be prefetched.

\lstset{escapeinside={(*@}{@*)}}
\begin{lstlisting}[language=C, caption=Repeat the attack 10000 times and count cache hits for each guessed value., label=list2, upquote=true]
uint16_t a[100],b[100];
initialize a[100] and b[100] with 100 random prime numbers;
uint16_t counts[256] = {0};

for t = 0 to 99{
    for k = 0 to 99{
        //Phase 1 & Phase 2: same as in Listing 1
        //Phase 3 (Attacker): Obtaining the secret
        for (uint8_t guess  = 0 to 255) {
            uint8_t mix = (guess*a[t]+b[t]) & 255;(*@\label{list2:sequence}@*)
            //Obtain the current clock cycle:
            uint64_t start = rdcycle();
            dummy = array2[mix * BlockSize];
            diff = rdcycle() - start;
            if(diff <= CACHE_HIT_THRESHOLD){
                counts[mix]++;
            }
        }
    }
}
\end{lstlisting}

The attack PoC is shown in Listing~\ref{list2}, where Phases 1 and 2 are the same as in Listing~\ref{list1}. In Phase 3, a disordered probing sequence ranging from 0 to 255 is generated by the way as shown in Line \ref{list2:sequence} in Listing~\ref{list2}, where a[t] and b[t] are predetermined with random prime numbers. The PoC of Spectre V1 attack shown in \cite{SpectreAttacksExploiting2019} used the same method to suppress the effects of the Stride Prefetcher.

\figurename~\ref{pic.denoise} shows the number of cache hits per 100-times attack for each guessed value, where the red dashed line is the average number of cache hits. When the number of clock cycles required to access the cache is less than CACHE\_HIT\_THRESHOLD , it is considered a cache hit. In this experiment, gem5 is configured as in Table~\ref{tab-gem5}, which has a CACHE\_HIT\_THRESHOLD of 35 cycles.

As illustrated in \figurename~\ref{subfig:denoise-pref} and \ref{subfig:denoise-dis}, while PREFENDER and DP lead to multiple cache hits at secret-unrelated entries in \textit{array2} due to prefetching, the attacker can still deduce that the secret value is `s' (ASCII value is 115) because it notably exhibits the highest cache hit count. This suggests that when an attacker is capable of repeatedly launching attacks and sampling, it can effectively filter out the noise introduced by prefetching and reliably infer the secret value.

As shown in \figurename~\ref{subfig:denoise-pcg}, in the case of PCG, the secret value `s' corresponds to a cache hit count of 62, which is around the average. This is due to the fact that PCG not only adds victim-irrelevant cache footprints as noise to the cache by utilizing prefetching, but also reduces the victim's cache footprints by prefetching back the lines that may be evicted by the victim's activity. This makes the victim's activity similar to the noise introduced by prefetching, which appears in the cache from time to time, and the attacker cannot reliably filter out the victim's access 
pattern through multiple attacks.

\subsection{Defense Against Enhanced Attackers}

In this subsection we discuss how the two types of enhanced cache-side channel attacks that can suppress prefetching are implemented and how PCG can provide an effective defense against them.

\mypara{Attacks with Pointer-Chasing Technique}
E. Tromer et al.~\cite{EfficientCacheAttacks2010}
first pointed out the effect of data prefetching on carrying out cache side-channel attacks and proposed the use of a random-order dual-chain structure to organize cache lines in eviction sets, employing a pointer-chasing approach for priming and probing~\cite{EfficientCacheAttacks2010}. This technique has been widely adopted in most of existing Prime+Probe attacks. 

The pointer-chasing technique is effective in suppressing stride prefetchers, because the attacker no longer installs and measures the eviction set in sequential, fixed increments. However, this technique cannot bypass PCG with active prefetching. Because no matter how the attacker accesses the eviction set to initialize or measure the cache set of interest, cache misses are incurred during this process. Thus the prefetching triggered by cache misses in PCG is insuppressible, which introduces a sufficient number of victim-irrelevant cache footprints to confuse the attacker.

\mypara{Prefetcher-Aware Prime+Probe Side-channel Attack (PAPP)} PAPP~\cite{PAPPPrefetcherAwarePrime2019} reverse-engineers prefetching policies and cache replacement policies to find out which cache line is replaced first when the victim accesses a particular cache set. Thus in the probe phase, the attacker only needs to measure one cache line rather than all the cache lines in the target cache set, to determine whether there is a victim access or not, drastically reducing the number of memory accesses and prefetcher noise.

PAPP is resistant to the interference of basic prefetchers, but it is difficult to work on the core that enables PCG. Firstly, PCG renders the reverse engineering of prefetching policies impossible. Because attackers cannot extract meaningful information related to the prefetching policy from random, noisy cache access patterns. Furthermore, even if PAPP is successful in identifying the cache line that will be evicted first, PCG makes it unfeasible to correlate the probing results with the victim's behavior. This is due to the fact that the attacker cannot distinguish whether the cache line miss is due to the victim access or due to PCG prefetching a new victim-irrelevant cache line. Similarly, the attacker cannot distinguish whether the cache line hit is due to the victim not accessing the target cache set or due to PCG prefetching the cache line that was evicted by the victim's activity.

\section{Evaluation}
\label{section:Evaluation}

\begin{table}[!t]
\footnotesize
\caption{gem5 parameters.}\label{tab-gem5}
\centering
\begin{threeparttable}
\begin{tabular}{c|c}
\Xhline{1px}
        \textbf{Parameter} & \textbf{Value} \\
  \Xhline{1px}
        \multicolumn{2}{c}{\textit{General}}\\
        \hline
        Processor type & RiscvO3CPU,  2-wide \\
         L1 ICache & \makecell{16KB, 4-way, 64 Sets}\\
        L1 DCache & \makecell{16KB, 4-way, 64 Sets}\\
        Prefetch Queue & 32 entries \\
        MSHR & 4 entries \\
        L2 Cache & 512KB, 16-way \\
        \hline
        \multicolumn{2}{c}{\textit{Next-Line Prefetcher (optional)}}\\
        \hline
        Degree & 4\\
        \hline
        \multicolumn{2}{c}{\textit{Stride Prefetcher (optional)}}\\
        \hline
        Degree & 4 \\
        Table entries & 64 entries\\
        Confidence counter bits & 3 bits \\
        Confidence threshold & 0.5 \\
        \hline
        \multicolumn{2}{c}{\textit{Signature Path Prefetcher (optional)}}\\
        \hline
        Signature table entries & 1024 entries \\
        Pattern table entries & 4096 entries \\
        Signature bits & 12 bits\\
        Confidence threshold & 0.5 \\
        
    \Xhline{1px}
\end{tabular}
\vspace{0.3em}
\end{threeparttable}
\end{table}
We have implemented PCG both in the gem5 simulator with a RISC-V core model and in the open-source RISC-V core BOOMv3\cite{SonicBOOM3rdGeneration2020}, to evaluate it in terms of security, performance, and hardware resource consumption overhead.

\subsection{Simulation Methodology}
The parameters used for implementing PCG in the cycle-accurate simulator gem5 are shown in Table~\ref{tab-gem5}. A 2GHz RISC-V single-core CPU with a 16KB L1 ICache, 16KB L1 DCache, and 512KB L2 Cache is the configuration for the baseline. For performance analysis, 21 benchmarks from SPEC17 suite~\cite{SPECCPU2017} are selected as the workloads of gem5. The first one billion instructions are skipped. Then the subsequent 10 million instructions are run with their IPCs (Instructions Per Cycle) recorded. 

We also implement PCG in BOOMv3, an open-source parameterizable high-performance out-of-order RISC-V processor. The original BOOMv3 is selected as the unsafe baseline. We use Chipyard v1.8.0 to generate RTL code, and then run the simulations through Verilator v4.210. By running Vivado, a Xilinx software suite for hardware design synthesis and analysis, the original BOOMv3 and security-enhanced BOOMv3 with PCG have been successfully synthesized with their on-board resource consumption measured.

\subsection{Performance Evaluation}
\label{subsection:Evaluation-Preformance}

\begin{figure}[!t]
  \centering
  \includegraphics[width=\linewidth]{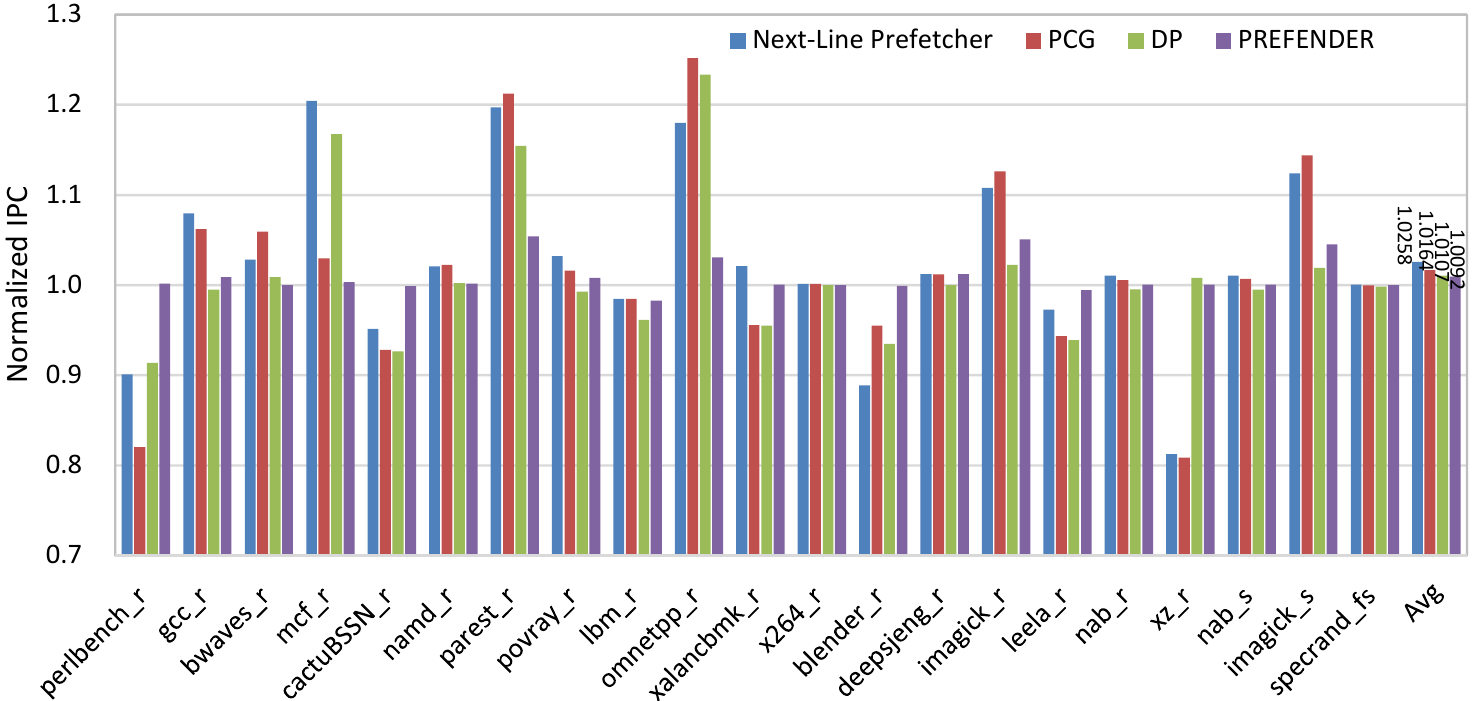}
  \caption{ {Performance improvement evaluation.}}
  \label{ipc}
\end{figure}

We implement the Next-Line Prefetcher, PCG, DP and PREFENDER in L1 Dcache based on the baseline. \figurename~\ref{ipc} shows the comparison of the normalized IPC among them, all running the 21 benchmarks selected from SPEC17. The Next-Line Prefetcher gains an average of 2.58\% performance improvement. PCG exhibits an average performance improvement of 1.64\%, DP shows a 1.07\% improvement, and PREFENDER shows a 0.92\% improvement. 

Overall, PCG can maintain or even improve performance while ensuring security. It employs prefetching to disrupt attackers. If prefetching is executed correctly and timely, it can lead to performance improvements. PCG shows the most noticeable performance enhancements for {\tt parest\_r} and {\tt omnetpp\_r}, with normalized IPC increasing to 1.21 and 1.25, respectively. Both of them have a large number of memory access operations~\cite{hassan2021reusable}, which can benefit from advance prefetching. However, the cache lines prefetched by PCG may not always be useful, which may potentially occupy cache resources and cause performance degradation. An example is {\tt prelbench\_r}, the normalized IPC of which approximately decreases to 0.82.

\mypara{PCG versus the Basic Prefetcher} {To be fair, we compare PCG with the Next-Line Prefetcher with the same prefetch degree of 4. The PCG performance degradation compared to the Next-Line Prefetcher is due to the PCG disrupting the order of prefetching as well as changing the original cache replacement policy. This is a sensible performance sacrifice to improve security.}

\mypara{PCG versus DP}
DP is an earlier innovative exploration of prefetching for mitigating cache side-channel attacks. It introduces a random prefetching and set-balance policy to extend basic prefetchers. For most benchmarks, PCG and DP exhibit similar performance, e.g., both result in performance reductions for {\tt perlbench\_r} and {\tt cactuBDDN\_r}, as their prefetching strategies are both similar to Next-Line prefetching.

Overall, PCG outperforms DP in terms of performance improvement. In order to achieve sufficient randomness and disruption, DP sets its prefetching degree to 10, which is too aggressive and lead to cache pollution. PCG sets the degree to 4 and triggers prefetching to add interfering noise on cache misses. Additionally, PCG adjusts the prefetching used to reduce the victim-related cache footprints based on the identification of abnormal cache sets. As a result, PCG will not cause excessive cache pollution for most benign programs.

\mypara{PCG versus PREFENDER}
PREFENDER can mitigate several cache side-channel attacks, including the attacks based on Flush+Reload, Evict+Reload, and Prime+Probe. Compared to PCG, it provides a wider defense range. However, it gains fewer performance improvement than PCG, because it only triggers prefetching when the {\tt load} instructions that could potentially be used in an attack are detected.

\subsection{Sensitivity analysis}
\label{subsection-sensi}
\begin{figure}[!t]
  \centering
  \includegraphics[width=\linewidth]{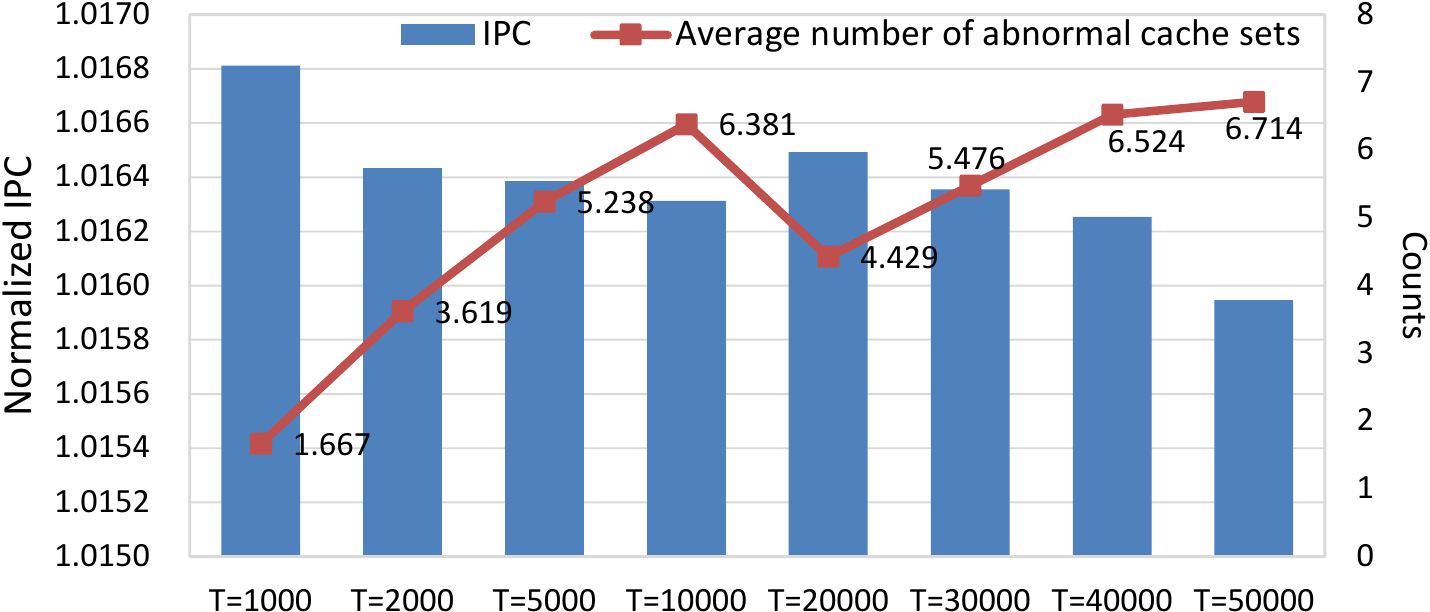}
  \caption{ {The impact of different reset periods.}}
  \label{threshold}
\end{figure}

 {For discussing the impact of the reset period of \textit{accessCounter} and \textit{dangerSet} on performance and security, we set $T$ to $(1000,2000,5000,10000,20000,30000,40000,50000)$. The corresponding normalized IPC and the average number of abnormal cache sets for running the 21 benchmarks are shown in \figurename ~\ref{threshold}. } 

 {Overall, as $T$ increases, the reset period of \textit{accessCounter} becomes longer, leading to more abnormal cache sets detected by AAM. Then OCM brings back more evicted cache lines and adjusts more prefetching lines into abnormal cache sets, enhancing security but resulting in a light performance decline. When changing $T$ from $10000$ to $20000$, the overall trend fluctuates. This is because some benchmarks, such as {\tt povray\_r}, access certain cache sets that belong to different reset periods, resulting in that these cache sets are no longer classified as abnormal.}

 {To analyze the security of PCG under different values of T, we conducted 1000 experiments using the method described in Section \ref{subsection:securityComparison} but random stall between Phase2 and Phase3 to obtain heat maps for different reset periods. Except for $T = 1000$, all other settings display noisy heat maps, indicating a relatively ideal level of security. Due to space limitations, they are not presented.}

\begin{figure}[!t]
  \centering
  \includegraphics[width=\linewidth]{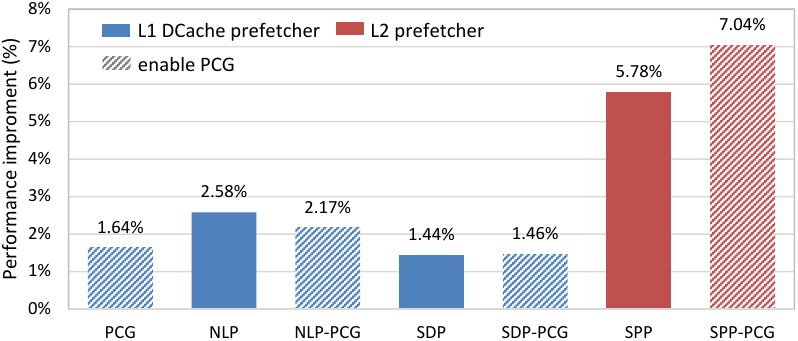}
  \caption{ {Performance improvement for PCG combined with basic prefetchers.}}
  \label{subfig:prefetcher}
\end{figure}

\subsection{Combine PCG with Basic Prefetchers}
\label{subsection:Evaluation-Miss}

PCG fundamentally operates as a prefetcher and can work combined with other basic prefetchers. We implement PCG in L1 DCache and combine it with Next-Line Prefetcher (NLP) or Stride Prefetcher (SDP). PCG is also implemented in L2 Cache and combined with Signature Path Prefetcher (SPP). So a total of 7 different prefetchers are evaluated. Using the configurations specified in Table~\ref{tab-gem5}, we show in Figure~\ref{subfig:prefetcher} the average IPC improvements normalized to the baseline by running 21 benchmarks. 


NLP and NLP-PCG exhibit improvements of 2.58\% and 2.17\%, respectively. SD and SD-PCG exhibit improvements of 1.44\% and 1.46\%, respectively. Based on the results of NLP and NLP-PCG, enabling PCG does not always guarantee a performance boost. This is due to the fact that NLP triggers prefetches with a degree of 4 for the cache access, and PCG triggers prefetches on cache miss. When they are combined, this may lead to an excessive amount of prefetched data that the core does not use, resulting in cache pollution and performance degradation.

When PCG is configured in the L2 Cache and combined with SPP, the performance improves from 5.78\% to 7.04\%, outperforming the results in L1 DCache. This is because the experiment employs an L2 Cache with a size of 512KB, which is significantly larger than the 16KB L1 DCache. Such a larger capacity can accommodate more prefetching requests while reducing the possibility of cache pollution.

\subsection{Validation based on BOOMv3}
\label{subsection:Evaluation-RTL}
The parameters of BOOMv3 are shown in Table~\ref{tab-boom}. The RTL code is then generated for the core by using Chipyard, which is an open-source platform for rapid prototyping and customized chip design. Based on the generated RTL, we run Evict+Reload attack similar to Listing~\ref{list1} through the Verilator.  {In the attack, the secret value is 115 (notated as ASCII code `s'), and the secret value is used to index a 256-item auxiliary array. We evaluated the average access latency of 1000 times to each auxiliary array item.}

\begin{figure}[!t]
  \centering
  \subfloat[For the BOOMv3 without PCG]{\includegraphics[width=0.95\linewidth]{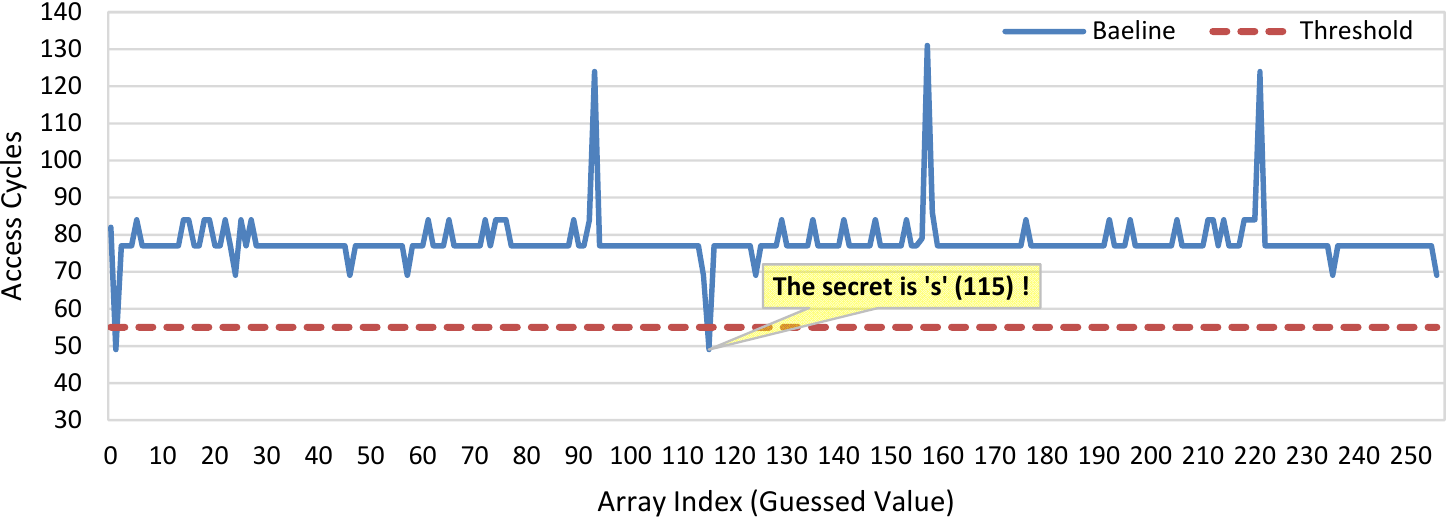}\label{subfig:log-nopcg}}
  
  \subfloat[For the BOOMv3 with PCG]{\includegraphics[width=0.95\linewidth]{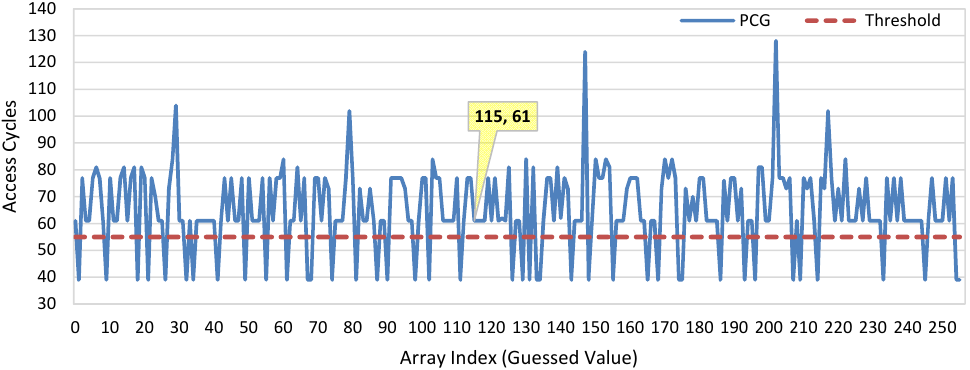}\label{subfig:log-pcg}}
  \caption{The logs of Spectre V1 attack on the BOOMv3 with and without PCG enabled.}
  \label{pic.attack_log}
\end{figure}

 {The results of the Evict+Reload attack running in the BOOMv3 without and with PCG enabled are shown in \figurename~\ref{pic.attack_log}. The red dashed line is the cache hit/miss threshold, which in BOOMv3 is 55 cycles. Without PCG, as shown in \figurename~\ref{subfig:log-nopcg}, the $115^{th}$ item in array with access cycles less than the threshold is shown as a cache hit, so the attacker can easily infer that the secret character is `s'. With PCG deployed, there are multiple cache hits excluding the $115^{th}$ item, as shown in \figurename~\ref{subfig:log-pcg}. Consequently, it becomes arduous for the attacker to determine the value of the secret character.}

\begin{table}[!t]
\footnotesize
\caption{BOOMv3 Parameters.\label{tab-boom}}
\centering
\begin{threeparttable}

\begin{tabular}{c|c}
\Xhline{1px}
        \textbf{Parameter} & \textbf{Value} \\
  \Xhline{1px}
        Processor type & {MediumBoom, 2-decode, 4-issue} \\
        \hline
        Pipeline depth & 10 stages \\
         \hline
         L1 ICache & \makecell{16KB, 4-way, 64 Sets}\\
        \hline
        L1 DCache & \makecell{16KB, 4-way, 64Sets}\\
        \hline
        MSHR & 4 entries \\
        \hline
        L2 Cache & 512KB, 16-way \\
    \Xhline{1px}
\end{tabular}
\end{threeparttable}
\end{table}

\subsection{Hardware Cost Evaluation}
\label{subsection:Evaluation-cost}

By running Vivado, which is a Xilinx software suite for hardware design synthesis and analysis, the original BOOMv3 and the BOOMv3 with PCG enabled are successfully synthesized with their hardware resource consumption obtained. The resources mainly include look-up table (LUT), flip-flop (FF), block RAM (BRAM), and digital signal processor (DSP). As shown in Table~\ref{tab.hardware-cost}, PCG incurs 1.26\% and 0.27\% overhead on the consumption of LUT and FF, respectively, and no extra consumption of both BRAM and DSP, indicating a comparatively lightweight hardware cost of PCG. 

\begin{table}[!t]
\centering
\caption{Hardware resource consumption.}
\label{tab.hardware-cost}
\adjustbox{width=0.45\textwidth}{
\begin{tabular}{l|c|c||c}
\Xhline{1px}
 & \makecell{BOOMv3 without\\PCG} & \makecell{BOOMv3 with\\PCG} & \makecell{{Overhead} resulted\\from PCG} \\ \Xhline{1px}
\# of LUTs & 169,463 & 171,165 & \textbf{1.26\%} \\ \hline
\# of FFs & 93,994 & 94,247 & \textbf{0.27\%} \\ \hline
\# of BRAMs & 179 & 179 & \textbf{0.00\%} \\ \hline
\# of DSPs & 36 & 36 & \textbf{0.00\%} \\ \Xhline{1px}
\end{tabular}}
\end{table}

\section{Related Works}
\label{section:Related}

This Section provides a review of existing research on both general and prefetching-based cache side-channel mitigation. 

\subsection{Secure Cache Designs}

Several secure cache architectures, based on cache partitioning or randomized mapping techniques, have been proposed to mitigate conflict-based cache side-channel attacks. 	

 {}Partition-based cache architectures divide the cache into secure and non-secure regions, preventing cache set conflicts between attackers and victims. CATalyst~\cite{CATalyst2016} utilizes Intel's cache allocation technology and page coloring to partition and isolate the LLC. PLCache~\cite{PLCache2007} locks protected cache lines to ensure they are not replaced by unprotected cache lines. Such approaches significantly impact overall cache utilization and fairness, and their effectiveness relies on the classification of secure and non-secure applications in the operating system.

Randomized-based secure cache architectures work by randomizing the mapping of memory addresses to cache sets, thereby obfuscating the location of victim data within the cache, and breaking the observability of cache access patterns. NewCache~\cite{Newcache2016} achieves randomization by employing a temporary logical cache between memory addresses and actual cache lines. MIRAGE~\cite{MIRAGE2021} has designed a practical fully associative cache that randomly selects candidate lines for replacement from all cache lines. However, it is difficult to implement and use these new cache designs due to their high complexity. In addition, the effectiveness of some programs related to the cache mapping strategy, such as prefetching, will be affected if the remapping of the cache is ignored.

\subsection{Prefetching-based Countermeasures}
Data prefetching was originally proposed to further reduce cache miss rate and improve system performance. The content brought into the cache by the data prefetcher becomes a natural noise injected into cache side channels~\cite{EfficientCacheAttacks2010}.

Some researchers have also been focusing on enhancing prefetchers to mitigate cache side- or covert-channel attacks. FLUSH+PREFETCH~\cite{FlushPrefetch2020} proposes the utilization of independent threads with {\tt clflush} and {\tt prefetch} instructions to randomly access the memory of secure applications. This approach requires program developers to comprehend the attack principles and carefully choose the objects to flush and prefetch. Additionally, it is worth noting that some of existing open-source cores, such as BOOMv3, do not support {\tt clflush}- and {\tt prefetch}-like instructions.


DP~\cite{Disruptive2015} introduces two secure-enhanced extensions to standard prefetching policies: 1) Random Prefetching Policy, which randomizes the order and {degree} of prefetches each time, and 2) Set-Balancer Prefetching Policy, which aims to distribute the prefetched cache lines across all cache sets. PREFENDER~\cite{PREFENDER2022} designs a Data Scale Tracker (DST) and an Access Pattern Tracer (APT) to interfere with Phases 2 and 3 of cache side-channel attacks, respectively. DST tracks the computation history of load instruction to predict another memory block that might be accessed by the victim and prefetch it. APT utilizes buffers to associate addresses from the same load and calculates the minimum distance, \textit{DiffMin}, between these addresses. If the current address is \textit{paddr}, then the prefetch addresses are \textit{paddr+DiffMin} and \textit{paddr-DiffMin}. However, both DP and PREFENDER, in essence, introduce noise into the cache, which attackers can attempt to filter out through repeated experiments.

BITP~\cite{BITP2019} focuses on utilizing prefetching to address the across-core LLC conflict-based attacks. When an attacker evicts a cache line in LLC, it results in a back-invalidation-hit in the L2 cache. BITP is used to prefetch these back-invalidation lines and refill them into L2 and LLC before the attacker accesses/observes them. Our work, on the other hand, is dedicated to mitigating the conflict-based cache side-channel attacks on the same cores. It complements BITP and aims to provide cache protection countermeasures with low hardware and performance overhead.

\section{Conclusion}
\label{section:Conclusion}

This paper proposed PCG, a security-enhanced scheme based on prefetching to mitigate conflict-based cache side-channel attacks. PCG reduces victim-relevant cache footprints by prefetching those lines that might have been evicted by the victim back to the cache, while also ensuring  that those lines installed by the victim are evicted from the cache as early as possible. PCG also employs prefetching to add balanced and noisy victim-irrelevant cache footprints. Thus, it is difficult for attackers to extract victim activities from confused cache access patterns.

PCG has been fully implemented both in the gem5 simulator and the realistic RISC-V core BOOMv3. We demonstrated the security of PCG by simulating high-precision Evict+Reload attacks and discussed how PCG effectively defends against prefetch-aware enhanced attacks. Additionally, we illustrated the reasons why and the scenarios where previous works relying solely on prefetch-induced noise, such as PREFENDER and DP, are ineffective in countering conflict-based cache side-channel attacks.

The performance evaluation based on SPEC17 showed that PCG has a very slight impact on processor performance, which is even improved on average. Additionally, according to the evaluation based on BOOMv3 synthesized by Vivado, PCG also leads to a very low hardware resource consumption.

\bibliographystyle{IEEEtran}
\bibliography{references}

\vfill

\end{document}